# Active Nanophotonics

Alex Krasnok, *Member IEEE*, and Andrea Alù, *Fellow IEEE*

*Photonics Initiative, Advanced Science Research Center, City University of New York, New York 10031, USA*

[akrasnok@gc.cuny.edu, aalu@gc.cuny.edu](akrasnok@gc.cuny.edu, aalu@gc.cuny.edu)

## Abstract

Recent progress in nanofabrication has led to tremendous technological developments for devices that rely on the interaction of light with nanostructured matter. Nanophotonics has hence experienced a large surge of interest in recent years, from basic research to applied technology. For instance, the increased importance of ultralow-energy data processing at fast speeds has been encouraging the use of light for signal transport and processing. Energy demands and interaction time scales become smaller with the physical size of the nanostructures, hence nanophotonics opens important opportunities for integrating a large number of devices that can generate, control, modulate, sense and process light signals at ultrafast speeds and below femtojoule/bit energy levels. However, losses and diffraction pose fundamental challenges to the fundamental ability of nanophotonic structures to efficiently confine light in smaller and smaller volumes. In this framework, active nanophotonics, which combines the latest advances in nanotechnology with gain materials, has recently become a vital area of optics research, both from the physics, material science, and engineering standpoint. In this paper, we review recent efforts in enabling active nanodevices for lasing and optical sources, loss compensation, and to realize new optical functionalities, like $\mathcal{PT}$-symmetry, exceptional points and nontrivial lasing based on suitably engineered distributions of gain and loss in nanostructures.

# I. Introduction

Nanophotonics has been experiencing a fast development in recent years, triggered by tremendous achievements in material science and nanofabrication[1]. This development led to advances in various vital applications, including microscopy [2]–[4], sensing [5]–[10], imaging [11], medicine [12], [13], light sources [14]–[25], and functional devices [26], [27]. These applications rely on the optical interactions of matter at the nanoscale. In the early stages of this field of research, the main driving force has been the discovery of plasmonic resonances in metallic structures, i.e., resonant enhancement of correlated oscillations of electrons arising at certain frequencies defined by material and geometrical parameters. These resonances are associated with small mode volumes, allowing strong enhancement of light-matter interactions near the metal surface [28]–[35]. As a result, metallic nanostructures provide new opportunities to generate concentrated bright hotspots, overcoming the diffraction limit and enabling microscopy and spectroscopy at extremely small, even up to the single-molecule, dimensions. Such fascinating optical properties have made plasmonics very promising for photonic applications [29], [36], [37]. Following developments in plasmonics have become an incentive for introducing new concepts, including metamaterials and metasurfaces [38], with unprecedented abilities to control light.

However, strong light localization in plasmonic materials is achieved at the price of increasing energy dissipation, which comes hand in hand with enhanced light-matter interactions [39]. The energy decay rate in metals is defined by electron-phonon and electron-electron scattering and is around $\gamma_{res} \approx 10^{13} - 10^{14}$ s$^{-1}$. This leads to fast plasmon dephasing and the conversion of energy into heat. As a result, the propagation length [$l_{spp} = 1/(2\alpha)$, with $\alpha$ being the field attenuation coefficient] of the surface plasmon-polaritons (SPPs) in noble metals (Au, Ag) is limited to several tens of micrometers in the near-IR ($\alpha \approx 10^2$ cm$^{-1}$) region, and it rapidly decreases to hundreds of nanometers in the visible range ($\alpha \approx 10^3 - 10^5$ cm$^{-1}$). In turn, the $Q$-factor of a plasmonic resonance [$Q_{res} = |\text{Re}(\omega_0)/2\text{Im}(\omega_0)| = \text{Re}(\omega_0)/2\gamma_{res}$, here $\omega_0$ is the (complex) resonance frequency, $\gamma_{res}$ is the total decay rate of the resonant mode] in noble metal nanoparticles does not exceed a few tens in the visible range. The search for new plasmonic materials [29], such as highly doped semiconductors and graphene led to some improvements, but not groundbreaking changes in this paradigm. In [40] it has been theoretically suggested that, by increasing the atom spacing by ~2, would significantly reduce material loss.

Another way to get around the significant loss in plasmonic materials is to replace metallic components of nanophotonic devices by high-index dielectric components, waveguides and nanoparticles (NPs), made of, e.g., silicon (Si), germanium (Ge), GaP [41]–[53], or halide perovskites [54] (hereinafter we define as "dielectric" a material with small conductivity at optical frequencies, i.e., this definition includes semiconductors). Light confinement in these structures relies on strong refractive index contrast. For example, the use of Si waveguides has established the entire field of silicon photonics [55]–[57], where field attenuation $\alpha < 10$ cm$^{-1}$ in the IR range is routinely realized. In the case of localized resonances, it has been demonstrated that low-loss high-index dielectric NPs and nanostructures provide lower dissipative losses along with a moderate light localization at the nanoscale. For example, nanodevices made of Si and other dielectrics demonstrate near unity scattering efficiency in the visible and near-IR [58], [59]. In addition, dielectric NPs have a number of other peculiarities, which are not typical for their plasmonic counterparts, including resonant magnetic optical response [47], [60], strong Raman scattering [43], [45], [52], [61], optical gain [54], [62] and possibility of integration with quantum dots (QDs) [63].

Although dielectric structures usually have significantly lower material loss, they are fundamentally limited by diffraction, given that their refractive index is limited to a few units. Hence, they cannot replace plasmonic nanostructures in applications relying on small-mode volumes and large electric field enhancement, such as surface-enhanced Raman scattering (SERS), strong coupling at the molecular scale, hyperbolic metasurfaces and metamaterials. Dielectric nanostructures still manifest some loss, which typically increases with larger operation frequencies [45]. Because of large radiation losses associated with their size, the *Q*-factor of low-loss dielectric NPs is mostly restricted to several tens [45] ($\gamma_{res} \approx 10^{13}$ s$^{-1}$), which can be tailored through the NP shape and composition and can be increased further in the case of quasi bound states in the continuum (BICs) [64]–[66].

In general, nanophotonic losses can be divided into two types: material and radiative ones. While different processes of energy dissipation (e.g., generation of phonons) are related to *material losses*, *radiative losses* account for electromagnetic energy going out from the localized volume of the nanostructure towards the far-field. Although progress in the fabrication of nanostructures has allowed reducing some material losses, such as those associated with roughness and material imperfections [67], there is a fundamental material loss (often called *intrinsic*) that cannot be suppressed by further developing of fabrication techniques [68], [69]. Radiative losses can be alleviated with improved fabrication techniques, and tailoring of nonscattering (cloaked) states or nonradiating field distributions, such as BICs [70]. In general, these nonradiative states can be achieved with specific nanostructure designs, e.g., utilizing zero-index materials [71], or by simultaneous excitation of two or more modes destructively interfering in the far-field. While radiative losses may be useful for scattering control or light amplification, they fundamentally hinder applications such as lasers, sensing and strong coupling regime.

The implementation of active materials in nanophotonics has recently appeared as a promising approach for loss compensation. In its beginning, this approach has been successfully implemented for loss compensation in nanoparticles, plasmonic waveguides, and metamaterials [39], [72], [73]. Later, new functionalities like lasing and signal amplification have also been demonstrated. Among them, nanolasers and *spasers* (acronym for "Surface Plasmon Amplifier by Stimulated Emission of Radiation") have attracted particular interest, because they allow pushing the concept of coherent stimulated emission down to the diffraction limit and beyond [74]. The concept of a spaser [75]–[81] has been initially suggested for the amplification of localized surface plasmons (LSPs) oscillating in metal nanoparticles and was generalized to include traveling surface plasmon-polaritons (SPPs) [67], [82]–[84]. In addition to their importance for all-optical and optoelectronic data processing, today they have been also employed for sensing, super-resolution imaging and biological imaging [74]. Beyond their tiny dimensions, nanolasers can have an ultrafast response, caused by accelerated spontaneous emission (Purcell effect) into lasing modes, allowing emission of subpicosecond pulses of coherent light.

More recently, it has been also realized that active nanophotonics is an ideal platform for the implementation of various concepts in non-Hermitian physics, including $\mathcal{PT}$-symmetrical systems and exceptional points into the optical arena [85]–[87]. These ideas date back to the seminal works of Bender and Boettcher [88], who mathematically discovered that non-Hermitian Hamiltonians (i.e., $\hat{H} \neq \hat{H}^+$, where $\hat{H}$ is the Hamiltonian of a system) can support real-valued spectra if they stay invariant with respect to simultaneous parity ($\mathcal{P}$-) and time-reversal ($\mathcal{T}$-) operations. In optics, $\mathcal{PT}$-symmetry corresponds to balanced distributions of gain and loss: a non-magnetic optical structure is $\mathcal{PT}$-symmetric if the real value of permittivity is an even function of the spatial coordinates, while the imaginary part is odd, requiring the presence of material gain. This gain-loss balance leads to the existence of exceptional points (EPs) at the real frequency axis in the exact $\mathcal{PT}$-phase, where two or more eigenvalues and their corresponding eigenvectors coalesce [85]. The EPs exhibit an abrupt symmetry breaking phase transition, once a parameter controlling the degree of non-Hermiticity (gain-loss parameter) exceeds a certain threshold, resulting in complex-valued spectra. Although EPs can exist in passive systems, their presence in active $\mathcal{PT}$-symmetric systems makes them promising for single-mode laser and advanced sensor applications [85], [89], [90]. In addition, the combination of such $\mathcal{PT}$-symmetric structures with various novel and recently emerging optical effects, like coherent perfect absorption (CPA), caused the appearance of a zoo of novel and fascinating effects for nanophotonics, such as CPA-lasers, unidirectional invisibility, and more [64], [91]–[93].

In this paper, we review the current stage of research and recent efforts in the broad field of active nanophotonics. We start from the general description of available approaches and materials, Section (II). Next, in Section (III) we provide a theoretical introduction to the problems involved in active nanophotonic systems, including Purcell effect, $\beta$-factor, rate equations, and scattering matrix approach. Then in Section (IV) we discuss different possible scenarios offered by non-Hermitian systems (i.e., having complex-valued dielectric permittivity distributions), including loss compensation and amplification, lasing, stability and $\mathcal{PT}$-symmetry. Section (V) is dedicated to a more detailed analysis of the state-of-the-art of loss compensation and amplification in nanophotonics. Section (VI) is devoted to lasing and spasing effects in nanostructures of different dimensions and materials with a focus on recently proposed approaches and materials (2D TMDCs, perovskites). Lastly, in Section (VII) we discuss how suitable engineering of distributions of gain and loss in subwavelength nanostructures may cause new functionalities, like $\mathcal{PT}$-symmetry, exceptional points, and nontrivial lasing. In our Conclusions, we summarize the paper and provide an outlook on the future development of this promising and rapidly growing field.

The interested reader may refer to several recent good review papers published on active metamaterials [94]–[96], active plasmonics [39], [97], [98], loss compensation [39], [99], nanolasers [74], [75], [100]–[104], spasers [81] as well as on $\mathcal{PT}$-symmetry in photonics [85], [105], quantum [106] and topological photonics [107]. It worth to mention that sometimes the term "active" in nanophotonics is used to indicate tuning of material properties (like the refractive index in phase-change materials) to control or reconfigure plasmon propagation [108]–[113]. In this paper, we use the term "active" in the sense of being able to provide energy to the optical wave, i.e., containing optical gain. This gain can be provided with gain materials (QDs, dyes) or with nonlinear effects (parametric amplification), as we discuss below.

## II. Available approaches and materials

*Gain materials*. As mentioned above, there are two main approaches to achieve optical gain in active systems, *gain materials* and *parametric amplification* via nonlinear effects. Fig. 1(a) shows the typical scenario of an active nanophotonic system in which an active material surrounds a lossy resonator (shown in blue). This structure may be realized with a bulk gain material (chromophore, QDs, halide perovskites, etc) dispersed with metallic/dielectric nanoparticles or considering a metal/dielectric core covered by a gain material shell. Below we discuss a 4-level gain material, whose typical Jablonski diagram is shown in Fig. 1(a), but similar considerations can be applied to other gain materials. The electrons of the gain material at the ground state "0" can be excited to a higher state "3" by light excitation of frequency $\hbar\omega_p = (E_3 - E_0)$. This process is characterized by the *excitation (pumping) rate* $W_p$, which depends on frequency, light intensity, and the internal quantum efficiency (average number of excited carriers per photon). Within the dipole approximation, the excitation rate scales as $W_p \propto |\mathbf{d}_{30} \cdot \mathbf{E}_p(\omega_p)|^2$, where $\mathbf{E}_p$ is the pumping field and $\mathbf{d}_{30} = e<3|\mathbf{r}|0>$ is the dipole moment of the transition $3 \to 0$.

The electron in the excited state, in general, may experience *spontaneous emission*, *stimulated emission*, or *nonradiative energy transfer* to other states [114]. Spontaneous emission, by its nature, generates incoherent photons, e.g., photoluminescence (PL), and hinders amplification, being a source of noise. Stimulated emission, in turn, creates photons coherent with the incident light. The amplification regime can be achieved when the stimulated emission rate overcomes the loss rate in the *population inversion regime*. Although population inversion cannot be reached in a simple 2-level system (in the CW regime), one can use a more complicated energy level structure, like the one depicted in Fig. 1(a). In this 4-level system, the excitation relaxes nonradiatively from "3" to the metastable state "2", from which it can relax to the lower state "1" with radiation of a stimulated or spontaneous photon with frequency $\omega_{21}$. In this case, the population inversion regime and optical amplification can be reached for a higher pumping rate $W_p$. Another way of realizing gain materials is by injecting carriers through electrical pumping, which is popular in semiconductor lasers [115], [116].

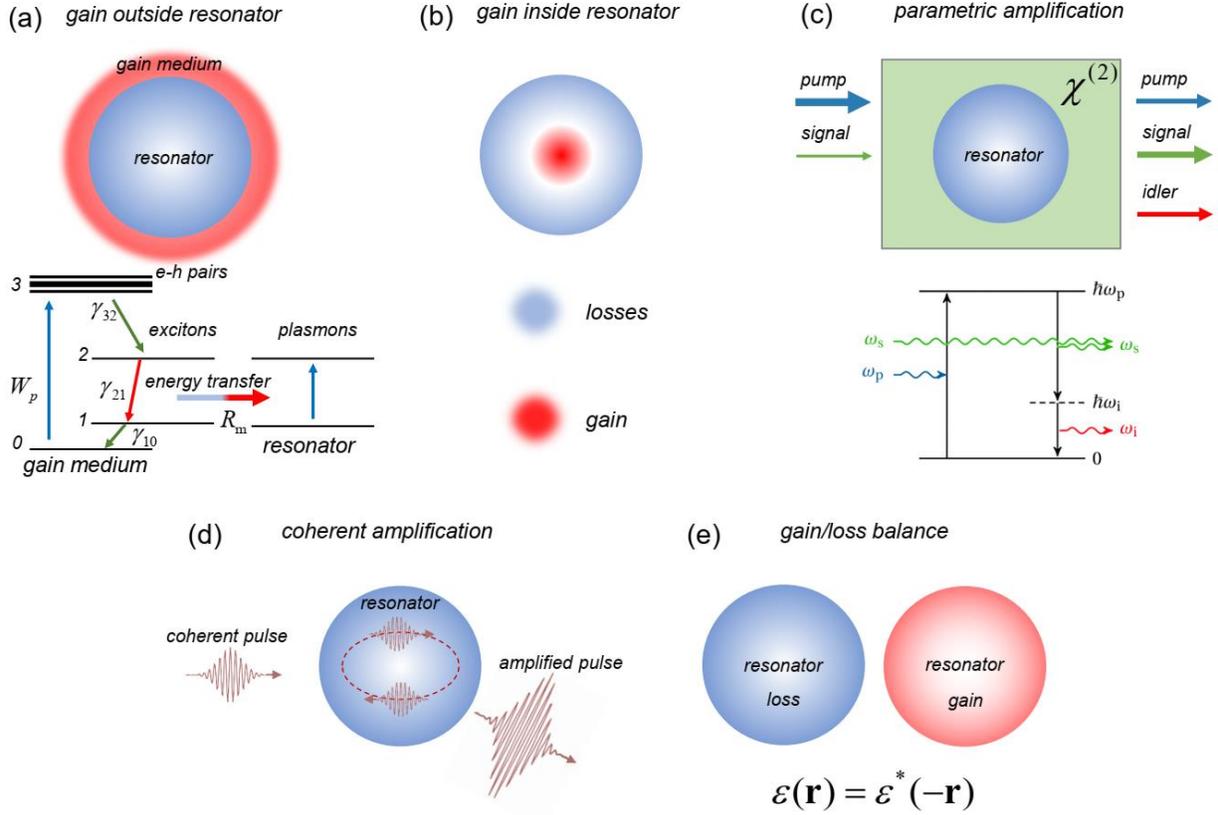

**Figure 1**. **Various approaches for active nanophotonics**. (a) A resonator (blue) is surrounded by a gain material (red). Below: the Jablonski diagram for molecular fluorescence excitation and energy transfer from gain material to the resonator. (b) Encapsulated gain material or intrinsic gain (*in situ* scenario) in a resonator. (c) Parametric amplification through nonlinear processes. (d) Coherent amplification. (e) Typical layout of $\mathcal{PT}$-symmetric nanostructure, in the form of two resonators with gain (red) and loss (blue).

The excited gain material can decay spontaneously in time with the decay rate $\gamma_{21,\text{rad}}^0$ in free space, which is usually measured by time-resolved PL spectroscopy [117]–[119] and nonradiatively with the rate $\gamma_{21,\text{non}}^0$. In the following, we denote by $\gamma_{21}^0$ the radiative decay rate, noting that the effect of nonradiative decay can be taken into account through the quantum yield $\eta$ factor, defined as $\eta = \gamma_{21,\text{rad}}^0 / (\gamma_{21,\text{rad}}^0 + \gamma_{21,\text{non}}^0)$ [26], [120]. Quantum yield, also known as quantum efficiency, depends on many factors, including material, environment and temperature. For reference, the quantum yield of gain materials typically used in photonics can reach ~1 – 10 % for transition metal dichalcogenides and carbon nanotubes [121], 30 – 50 % for bulk perovskites [122], 80% for fluorescent semiconducting polymers [123], and up to 90 – 100 % for perovskite and semiconductor quantum dots [124]–[126] even at room temperature. Further, the quantum yield can be significantly boosted via Purcell effect (radiative decay rate increasing), as discussed below [121], [127]–[129].

Next, we assume that there is only one channel of radiative energy decay; otherwise, the decay becomes multiexponential. In free space, the decay rate equals $\gamma_{21}^0 = \omega_{21}^3 |\mathbf{d}_{21}|^2 /(3\pi\varepsilon_0 \hbar c^3)$, assuming that the dipolar contribution dominates radiation due to the small size of the emitter, where $\varepsilon_0$ is the dielectric constant and $\mathbf{d}_{21} = e<2|\mathbf{r}|1>$ is the dipole moment of the transition $2 \to 1$ [130]. The decay rate $\gamma_{21}^0$ defines the decay time $\tau_{21} = 1/\gamma_{21}^0$, whose typical values lie between ~10 ns for QDs [131] and several 100 ns for dyes [132] and can be found in spectroscopy handbooks. Spontaneous photons can then induce the emission of stimulated photons giving rise to *amplification of spontaneous emission* (ASE), which has some similar properties (emission line narrowing, nonlinear input-output behavior) to lasing, sometimes complicating the classification of laser radiation [133].

At some level of population difference, stimulated emission overcomes intrinsic losses in the gain material, and the material starts to amplify the resonant electromagnetic fields (with simultaneous generation of spontaneous photons). In most cases, in *steady-state,* the gain material can be described by the complex-valued dielectric permittivity $\varepsilon_g = \varepsilon_g' + i\varepsilon_g''$, or the complex refractive index $n = n_0 + i\kappa$. In this paper, we use the time-harmonic convention $\exp(-i\omega t)$, therefore a gain material has $\kappa < 0$ ($\varepsilon_g'' < 0$) and a lossy material has $\kappa > 0$ ($\varepsilon_g'' > 0$) [134]. The parameter $\kappa$ is a coefficient representing the property of a material to coherently amplify ($\kappa < 0$) or to absorb ($\kappa > 0$) an electromagnetic field. It is more convenient to characterize a material gain by the so-called *material gain parameter* $g = 4\pi\kappa/\lambda$ [cm$^{-1}$], where $\lambda$ is the free-space wavelength, so that the light intensity $I$ changes with the propagation coordinate $z$ as $dI(z) = g(\omega)I(z)dz$. In general, the gain coefficient depends on the transition cross-section $\sigma_{21}$ and the population difference $N = N_2 - N_1$ of a gain material, $g = N\sigma_{21}$. We do not consider the case degenerate states. If $(N_2 - N_1) > 0$, then $g > 0$ and the material amplifies. In turn, the transition cross-section $\sigma_{21}$, in the dipole approximation, depends on the dipolar moment of the transition $\mathbf{d}_{21} = e < 2|\mathbf{r}|1>$ [130]. The material gain parameter required to achieve the lasing regime (see below) is called the *threshold gain coefficient* $g_{las}$. For example, in metal nanoparticle-based lasers, the value of the threshold gain coefficient depends on the material, shape and resonance frequency, and is in the order of ~ $10^3 - 10^4$ cm$^{-1}$ [135]–[137]. For loss compensation in Si waveguides a gain of ~ 1 – 100 cm$^{-1}$ is required, whereas plasmonic waveguides supporting SPP modes may require much larger gain.

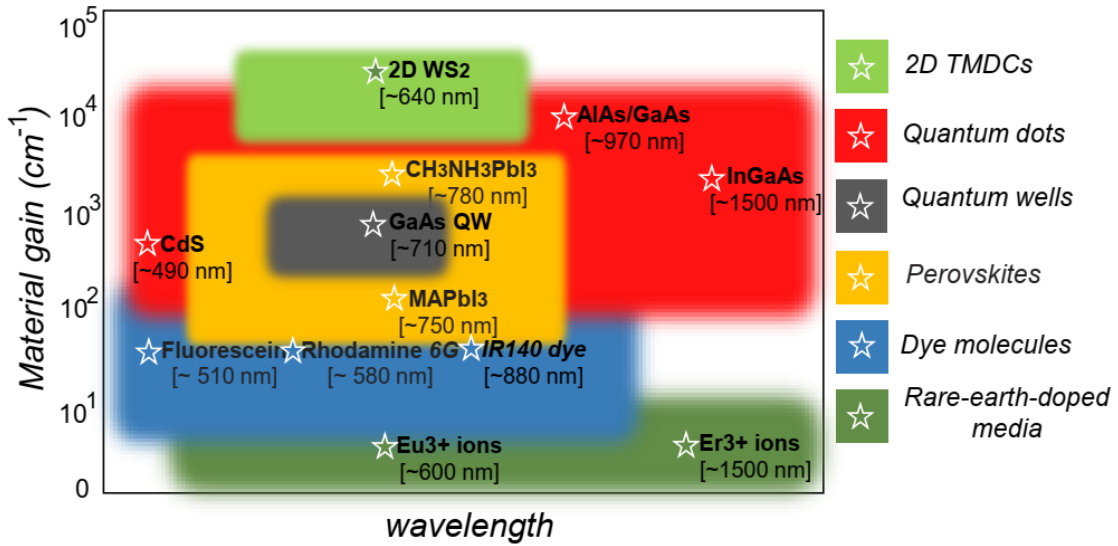

**Figure 2.** Achievable material gain parameter for different materials, including rare-earth-doped materials, dye molecules, perovskites, quantum wells, quantum dots, and 2D transition metal dichalcogenides (TMDC).

An important requirement to achieve loss compensation and lasing is a significant overlap of the gain distribution with the modal profile of the structure, characterized by the confinement factor $\Gamma$. Physically, $\Gamma$ defines the portion of the entire gain medium that is coupled to a certain laser mode (for definition and expressions of this factor in various systems see, for example [138]). To achieve a large confinement factor, the gain material can be encapsulated inside a nanoresonator, Fig. 1(b), for example in the form of a core-shell structure with a metal shell [139]. Dielectric nanoparticles showed themselves as promising for this purpose because they allow incorporation of QDs inside the particle or even can demonstrate intrinsic gain [62], [63] that leads to a significant overlap factor. Important classes of such dielectric materials allowing optical gain and optical resonances being nanostructured are halide perovskites[54], [140]–[142] and metal-organic frameworks (MOFs) [143]. Such nanostructures make feasible the *in-situ* loss compensation and lasing without difficulties in their fabrication (no adjustments of a mode and a gain material needed).

It is useful to review the maximum value of material gain coefficient reported in the literature for some of the most used gain materials, as shown in Fig. 2. Rare-earth-doped materials (optical glasses) demonstrate the smallest gain parameter, typically in the range of few units of cm$^{-1}$. Although these values are sufficient to compensate for losses in optical fibers, they are completely insufficient for nanophotonic applications. Next, dye molecules have gain parameter of ~10-100 cm$^{-1}$ [54], [144]. These gain materials are actively used for loss compensation in SPPs [39]. Perovskites, materials attracting significant attention today for active nanophotonics, provide larger material gain ~$10^2$-$10^3$ cm$^{-1}$. Then, quantum dots demonstrate higher material gain of $10^2$-$10^4$ cm$^{-1}$[145]–[147]. Finally, 2D transition metal dichalcogenides (2D TMDC) demonstrate strong optical gain up to few $10^4$ cm$^{-1}$[148]. Thus, the required amount of material gain for nanophotonic applications can be achieved in dye molecules, perovskites, quantum wells, quantum dots, and 2D TMDC.

*Other gain approaches.* Although in this paper we focus on material gain, because of its importance and widespread use in nanophotonics, other approaches are also available. Another method to impart gain is optical parametric amplification (OPA), consisting in the use of material nonlinearity. If a system with a nonlinear response is irradiated with two fields, a strong pump field at the frequency $\omega_p$ and a weak probe signal at frequency $\omega_s$ [as shown in Fig. 1(c)], their interaction in the nonlinear material allows generation of an idler wave with frequency $\omega_i = \omega_p - \omega_s$, propagating in the same material. The same nonlinear process is allowed between idler and pump, $\omega_s = \omega_p - \omega_i$, leading to the energy transfer from pump to signal, realizing optical parametric amplification [149]. This approach has been successfully implemented for loss compensation in metamaterials [95] and nanostructures [150].

A related approach, consisting in the use of periodic modulation in time, is available, leading to time-Floquet systems in which wave propagation is described by partial differential equations with time-periodic coefficients [151], [152]. Time-Floquet systems have been used in electronics [153]–[155] and wave engineering [156]–[158], leading to several exciting phenomena, including optical gain [158].

Another approach, *coherent amplification,* can provide gain, based on pulse amplification by constructive interference inside a cavity [159], see Fig. 1(d). This approach consists in matching the repetition period of an external pulsed laser with the cavity round-trip time. The incoming pulses constructively interfere with the intracavity pulse, which leads to pulse amplification. The intracavity pulse is switched off when enough energy is achieved. In this scheme, the optical cavity is assumed to be passive, and amplification relies on external pumping from the laser [159], [160]. The related coherent approach has been recently proposed for loss compensation in plasmonics [161].

## III. Theoretical Background

***Purcell factor and $\beta$ factor.*** It is illustrative to consider basic parameters driving the system of a gain material coupled to a nanophotonic resonator, as sketched in Fig. 3. When the gain material is coupled to a resonator, like a metal (plasmonic resonances) or dielectric (Mie resonances) nanoparticle, its dynamics change as follows. Firstly, the excitation rate $W_p$ becomes proportional to the local field intensity $|\mathbf{E}_p|^2$ at the excitation frequency, which can be enhanced at the resonance significantly. In other words, the resonator can boost the excitation of the gain material so that the required population can be reached at a smaller excitation intensity. Otherwise, the gain medium can be pumped by a current as it often takes place in semiconductor lasers[116].

Secondly, when the gain material is coupled to a resonator, its decay rate is also modified. This change affects both spontaneous and stimulated emission equally. In this coupled system, in addition to the direct radiation decay rate $\gamma_{21}^0$ to free space (which also can be changed), the energy from the gain material goes to the resonator (to all possible modes) radiatively or through Förster energy transfer with the rate $R_r$. This energy rate consists of spontaneous $R_r^{sp}$ and stimulated $R_r^{st}$ rates, $R_r = R_r^{sp} + R_r^{st}$. In general, the energy transfer rate to the resonator $R_r$ depends on the geometry of the system, the modal profile and its overlap with the gain material described by the confinement factor $\Gamma$. The value of gain times the confinement factor is called modal gain, $g_m = \Gamma g$. In general, it is defined via the spatial integral over the gain-modal distribution [162].

Thus, the modified total decay rate of the gain material becomes $\gamma_{21} = \gamma_{21}^0 + R_r = \gamma_{21}^0 + R_r^{sp} + R_r^{st}$. The value of $\gamma_{21}$ defines the decay time ($\tau_{21} = 1/\gamma_{21}$) required by the system to get to thermodynamic equilibrium. In the regime of stationary pumping, all the energy that gets to the resonator decays with the total decay rate of the resonator, $\gamma_{res}$, which is the sum of dissipation rate $\gamma_{nr}^r$ and radiation rate $\gamma_r^r$, i.e., $\gamma_{res} = \gamma_{nr}^r + \gamma_r^r$, Fig. 3.

The change of unperturbed decay rate ($\gamma_{21}^0$) induced by the presence of the resonator in the regime of no stimulated emission (i.e., $R_r^{st} = 0$, weak pumping) is characterized by the (total) *Purcell factor* [163]

$$F_{tot} \equiv \frac{\gamma_{21}}{\gamma_{21}^0} = 1 + \frac{R_r^{sp}}{\gamma_{21}^0}. \tag{1}$$

The Purcell factor numerically characterizes the *Purcell effect*, i.e., changing (reducing or increasing) of the decay rate of an emitter near a resonator [163], [164]. The Purcell factor can be found analytically [165]–[173] or numerically [169] for different resonators. Of special interest is the *single-mode* scenario, when the Purcell factor ($F_{tot}$) depends on the Q-factor and the effective volume ($V_{eff}$) as $\sim Q/V_{eff}$ and defined on resonance as [165], [174]:

$$F_{tot} = \frac{3}{4\pi^2} \frac{(\lambda/n_h)^3}{V_{eff}} \left( \frac{1}{Q_{res}} + \frac{1}{Q_{emit}} \right)^{-1}, \tag{2}$$

where $Q = (1/Q_{res} + 1/Q_{emit})^{-1}$ is the total quality factor of the system, $Q_{emit}$ and $Q_{res}$ are Q-factors of the uncoupled emitter ($Q_{emit} = \omega_{21}/2\gamma_{21}^0$) and resonator ($Q_{res} = \omega_0/2\gamma_{res}$), respectively. There are two characteristic conditions: *bad resonator* ($Q_{res} \ll Q_{emit}$) and *good resonator* ($Q_{res} \gg Q_{emit}$) limits. Since the typical emission spectrum of a gain material (dyes, QDs) is usually ~10 nm and $Q_{res} \sim 10^1 - 10^2$, in nanophotonics we usually deal with the bad resonator case and, therefore, the Purcell factor $F_{tot}$ can be defined as $Q_{res}/V_{eff}$ [175]. In turn, in the good resonator limit, i.e., high-Q cavity (disk cavity, defect in a photonic crystal, etc.) and the emitter at room temperature (fast dephasing), the Purcell factor is defined by $Q_{emit}$.

Only a portion of the total energy transfer rate $R_r$ couples to the *specific mode*, $R_m$. This is the only useful portion for amplification and lasing, and it also consists of spontaneous and stimulated parts, $R_m = R_m^{sp} + R_m^{st}$. This portion of the total energy transferred to the mode gets dissipated and radiated with the decay rate of the *resonator mode*, $\Gamma_m$, Fig. 3. The efficiency of coupling to the specific mode is defined by the $\beta$-factor

$$\beta = \frac{R_m^{sp}}{\gamma_{21}} = \frac{R_m^{sp}}{\gamma_{21}^0 + R_r^{sp}}, \tag{3}$$

which describes the fraction of spontaneously emitted photons radiated to the specific mode of interest. The same quantity defines the fraction of stimulated photons that get to the specific mode and hence enters the rate equations in the same manner (see below). The closer $\beta$ is to 1, the larger portion of energy couples directly to the mode, resulting in lower amplification and laser thresholds, as discussed below. For example, if the $\beta$-factor equals 0.5, only half of the emitted photons, on average, couple to the specific mode and, hence, the gain material must emit two times more photons to achieve amplification or lasing.

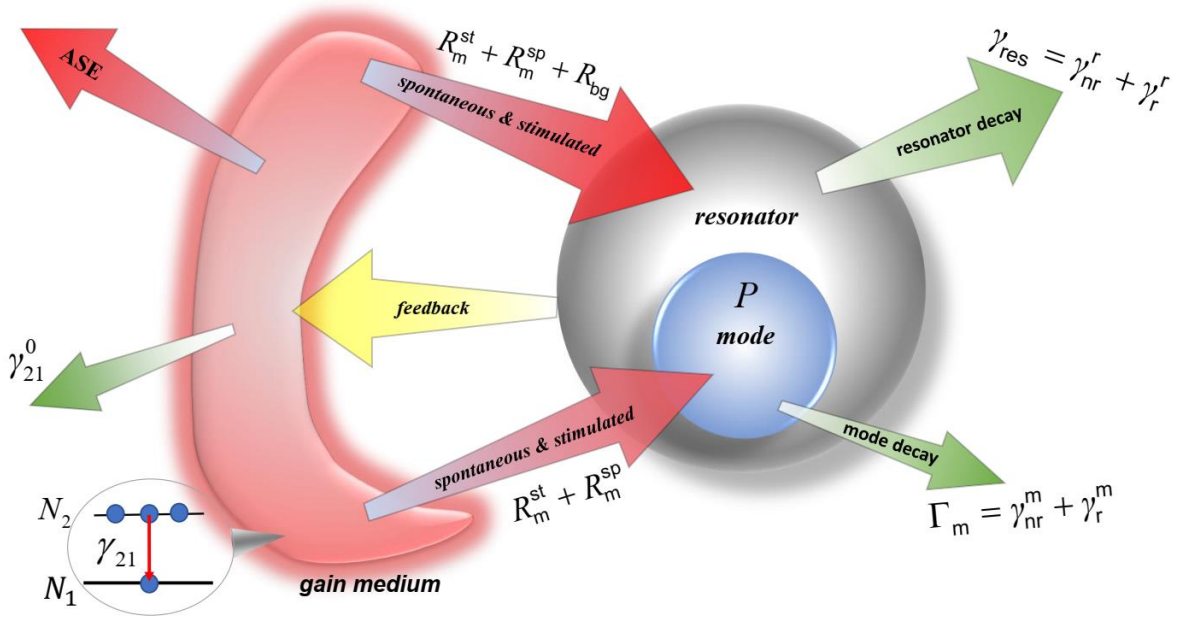

**Figure 3.** Illustration describing the rate equations. Here, ASE stands for amplified spontaneous emission, $\Gamma_m$ is the decay rate of the specific laser mode, $\gamma_{res}$ is the total decay rate of the entire

resonator, $\gamma_{21}^0$ is the gain medium decay rate in free space, $\gamma_{21}$ is the decay rate (of the spontaneous and stimulated emission) modified by the resonator, $R_r$ and $R_m$ are total energy transfer rate to the resonator and the portion that goes to the specific mode, respectively. The feedback is characterized by confinement factor $\Gamma$.

The $\beta$-factor is typically small in macroscopic lasers (e.g., typically $10^{-5}$ - $10^{-3}$ for VCSELs and edge-emitting lasers[116]) because of their multimode operation, resulting in a well pronounced amplified spontaneous emission (ASE) kink in the input-output characteristic[176]. For metal particle-based nanolasers (spaser), the value of $\beta$-factor significantly drops from units of $\sim 10^{-1}$ to $\sim 10^{-3}$ as a luminescent molecule gets closer to the particle surface [135]. This effect is known as *quenching* and it consists in the excitation of highly dissipative higher-order modes in the nanoparticle [177], [178]. In semiconductor-based nanolasers, the $\beta$-factor can reach much larger values, e.g., ~0.5 in Ref. [78]. Note that for large factors $\beta \to 1$, the kink in the input-output characteristic disappears, and the system demonstrates a *thresholdless* laser regime [179]–[183]. Such large values of $\beta$ require a more careful analysis of the laser characteristics and in particular a proper definition of the laser threshold [176], [184], as discussed below.

Dividing numerator and denominator in Eq. (3) by $\gamma_{21}^0$ we outline the relationship between $\beta$-factor and Purcell factor, $\beta = F_m / F_{tot}$, where $F_m = R_m^{sp} / \gamma_{21}^0$ defines the acceleration of emission (both dissipative and radiative) to the specific mode[185]. The remaining part of the total Purcell factor $(F_{tot} - F_m)$ corresponds to the change of energy decay rate to all other modes. Thus, in order to make the laser threshold smaller, one has to increase $\beta$ by reducing the number of modes, increasing the mode Purcell factor $F_m$ as much as possible, and simultaneously reducing $(F_{tot} - F_m)$.

*Rate equations.* Taking into account all the processes discussed above, the dynamics of the coupled gain-resonator system can be described (weak-coupling regime) by so-called rate equations [186]–[188]:

$$\frac{dN_2}{dt} = W_p - \gamma^0_{21,\text{non}} N_2 - \beta^{-1} R^{\text{st}}_{\text{m}} - \beta^{-1} R^{\text{sp}}_{\text{m}} + F_N(t),$$

$$\frac{dP}{dt} = -D_{\text{m}} + R^{\text{st}}_{\text{m}} + R^{\text{sp}}_{\text{m}} + F_P(t),$$

(4)

where $N_2$ is the electronic excited state population, $P$ is the photon mode population, $R^{\text{sp}}_{\text{m}} = \beta \gamma_{21} N_2$ and $R^{\text{st}}_{\text{m}} = \beta \gamma_{21} P(N_2 - N_1)$ are spontaneous and stimulated emission rates to the specific mode. Note that these expressions for $R^{\text{st}}_{\text{m}}$ and $R^{\text{sp}}_{\text{m}}$ are valid only for point emitters with the same absorption and emission cross-sections. The quantity $D_{\text{m}} = \Gamma_{\text{m}} P$ defines the mode damping, which consists of radiative and nonradiative (dissipative) parts, $D_{\text{m}} = D^{\text{r}}_{\text{m}} + D^{\text{nr}}_{\text{m}}$. Note that both spontaneous and stimulated rates are proportional to the total decay rate, $\gamma_{21}$, and hence can be enhanced by the Purcell effect. The values of $F_N(t)$ and $F_P(t)$ are stochastic Langevin forces, which define the noise and statistics in the system [186].

It follows from these equations that amplification may be achieved when $N_2 > N_1$, which corresponds to the population inversion regime. In turn, the lasing threshold in the steady-state is reached when *the number of stimulated photons in the mode overcomes the mode damping ($D_{\text{m}}$) and the number of spontaneous ones*, i.e., when $\gamma_{21} \beta P(N_2 - N_1) \geq \gamma_{21} \beta N_2$ [176]. Another criterium of the lasing regime in the steady-state is the $P \geq 1$, that is the so-called *quantum threshold condition* [176]. Note that the presence of Langevin forces compels us to use the arguments of radiation statistics to determine the regime of laser generation, see Section (IV) for more discussions.

*Dielectric permittivity of gain material*. The easiest way to incorporate gain into the analysis of a nanophotonic system is to assume that the material is described by the frequency-independent gain parameter, $\kappa < 0$ ($\varepsilon''_g < 0$). While this approach does not comply with Kramers-Kronig relations [134], it can still be used in the vicinity of a resonance [139], [189], [190] for the sake of simplicity or estimation of the gain parameter from an experiment [191].

However, the gain is inherently dispersive, and it is typically narrowband in nanophotonic systems. While for a rigorous analysis of the spatiotemporal dynamics of an active system the *rate equations* (4) should be used [175], [186], [192], the dispersive dielectric permittivity approach can be utilized for the steady-state analysis. As shown in Refs. [193]–[195], the dielectric permittivity of a 4-level active material (e.g., a chromophore) coupled to a nanoresonator [Fig. 1(a)] in the assumption of fast polarizations dephasing (entropy relaxation rate, $T_2^{-1}$) rate $\gamma_L$ (i.e., $\gamma_L \gg \gamma_{ij}$) and fast population relaxation (energy relaxation rate, $T_1^{-1}$) from levels 3 and 1 $(\gamma_{32}, \gamma_{10}) \gg (\Gamma_{res}, W_p, \gamma_{21})$, can be described by the gain permittivity

$$\varepsilon_g(\omega) = \varepsilon_h - G \frac{(\omega_{21} - \omega + i\gamma_L/2)(\gamma_L/2)}{(\omega_{21} - \omega)^2 + (\gamma_L/2)^2 (1 + |\mathbf{E}_r|^2 / E_{sat}^2)}. \tag{5}$$

This model works well in the range not very far from the frequency of the resonator mode $\omega_0$, $\omega_0 - \omega_{12} < \gamma_L$, and takes into account the gain saturation effect, restricting the maximum amount of gain that can be experimentally achieved. Here, $\varepsilon_h$ stands for the permittivity of the host material. The emission $2 \to 1$ of the gain molecules is described by a Lorentzian with amplitude $G$ (unsaturated gain level) and width $\gamma_L$. The dimensionless unsaturated gain level is associated with the pumping rate $W_p$ [s$^{-1}$], $G = \frac{c}{\omega} \sqrt{\varepsilon_h} N \sigma_{21} \frac{W_p}{\gamma_{21} + W_p}$, where $N$ is the density of molecules and $\sigma_{21}$ is their (orientation-averaged) emission cross-section in the bulk host material. The pumping rate $W_p$ depends on the (local) pumping field $\mathbf{E}_p$ and the gain material bulk absorption cross-section, $W_p = \frac{c \varepsilon_0 \sqrt{\varepsilon_h} \sigma_{30}}{2 \hbar \omega_p} |\mathbf{E}_p|^2$, where $\omega_p = (E_3 - E_1)/\hbar$ is the pumping frequency. In turn, the field strength $E_{sat}$ of saturation can be found as $E_{sat}^2 = \frac{2\hbar \omega_s (\gamma_{21} + W_p)}{c \varepsilon_0 \sqrt{\varepsilon_h} \sigma_{21}}$.

This model, in contrast to other analytical models [196]–[199], takes into account the nonlinear saturation phenomena and has been successfully applied to the analysis of optically pumped spasers and nanolasers [193], [195]. The model can also be generalized to more complex structures, like waveguides, metasurfaces, and metamaterials. Note that when the resonator field ($|\mathbf{E}_r|$) is small enough, a linearized model can be used [200]–[205].

***S-matrix description***. The scattering matrix ($\hat{S}$) connects input ($a_i$) and output ($b_i$) signal amplitudes of a linear system, $\mathbf{b} = \hat{S}(\omega)\mathbf{a}$ [64], [206]–[208]. The amplitudes are normalized such that $|a_i|^2$ and $|b_i|^2$ correspond to the energy of incoming and outgoing waves in channel $i$. For example, in a single port system, the scattering matrix coincides with the reflection coefficient $r$. In a two-port system, the scattering matrix is $\hat{S} = \begin{pmatrix} r_{11} & t_{12} \\ t_{21} & r_{22} \end{pmatrix}$, where $r_{ii}$ and $t_{ij}$ stand for the reflection and transmission coefficients of each port. The analysis of the S-matrix eigenvalues in the complex frequency plane demonstrates the existence of two types of point-like singularities, *zeros* and *poles* [64]. The poles correspond to self-sustained solutions (system eigenmodes) [64], [209], [210], whereas the zeros correspond to solutions without outgoing fields (perfect absorption). The analytical properties of the $\hat{S}$-matrix ensure that the knowledge of poles and zeros allows predicting the response of a system over the entire complex frequency plane [211]–[213]. In the next Section we use the S-matrix analysis for description of different scenarios in gain/loss systems.

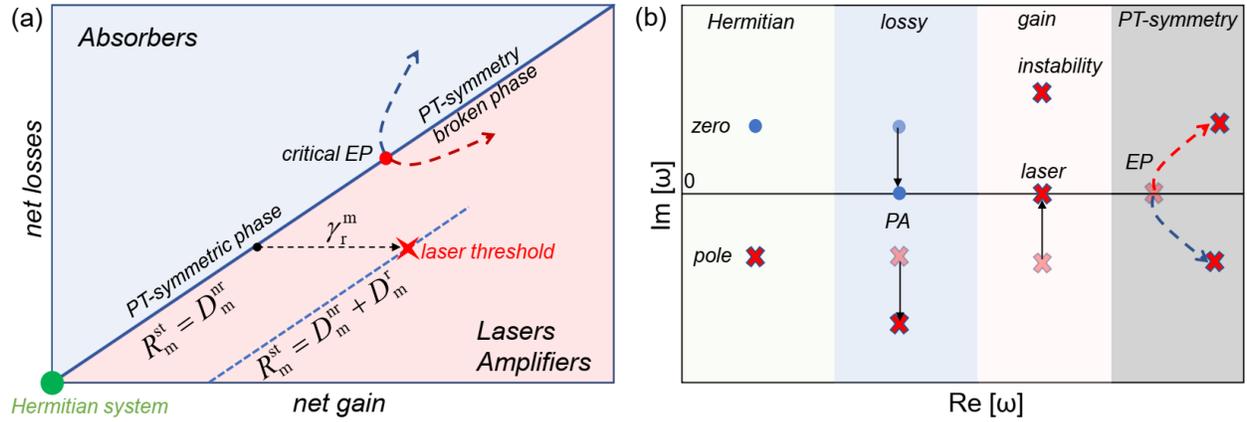

**Figure 4.** (a) Possible scenarios for an active system with gain and loss. (b) Poles and zeros analysis of scattering matrix in various non-Hermitian and active systems, including Hermitian structure, lossy system, gain system and $\mathcal{PT}$-symmetric structure.

## IV. Different scenarios in gain/loss systems

In this section, we discuss various scenarios in nanophotonics when gain/loss is present in a nanostructure, Fig. 4(a). We describe these different scenarios within the framework of the scattering matrix poles and zeros, as sketched in Fig. 4(b). Figure 4(a) represents a loss/gain diagram, in which the vertical and horizontal axes correspond to the net loss and gain in a system.

The green circle shows the *Hermitian regime*, in which there is no gain or loss[64]. In the lossless Hermitian system, poles $\omega_{pl,n}$ and zeros $\omega_{zr,n}$ of $\hat{S}$-matrix occur in complex conjugate pairs, i.e., $\omega_{zr,n} = \omega_{pl,n}^*$ and restricted to specific parts of the complex frequency plane. We use the time convention $\exp(-i\omega t)$, hence the poles (zeros) lie in the lower (upper) half-plane, Fig. 4(b, green area). Next, the thick blue solid line in Fig. 4(a) shows the regime in which the material loss is balanced by gain, i.e. when net losses are equal to a net gain in the system. If the system has radiative losses, this line of balance gets shifted to the right side so that the system needs more gain to compensate losses entirely. All scenarios below this line (red sector) correspond to amplification regimes, while the regimes located above the line (blue sector) are lossy ones when the net losses exceed the gain. All absorbing regimes are located in the blue sector. In the complex plane, adding loss leads to zeros (poles) getting closer (farther) to the real axis, Fig. 4b, (blue area). When a zero touches the real frequency axis, the system becomes *perfectly absorbing* (PA) at that frequency. An exciting realization of PA is the so-called *coherent perfect absorption* (CPA), or *time-reversed lasing*, feasible in two or more port systems[64], [93], [214]. This regime manifests itself as a perfect absorption of all energy coming from all ports when excited *coherently* with specific phase and amplitude relations [93], [215]. The generalization of this effet to low-loss systems realizes *virtual perfect absorption* (VPA) [214].

For a given system with some amount of material and radiation losses, the increase of gain allows *loss compensation*. In the complex plane, it corresponds to pushing the pole closer to the real axis, Fig. 4(b, red area). Once the material loss is compensated, i.e., when $R_m^{st} = D_m^{nr}$, the amplification regime is achieved, and a signal passing through the system grows. We remind that the quantity $R_m^{st}$ denotes the rate of stimulated energy transfer to the specific mode. In an experiment, in systems with small $\beta$-factor, this amplification regime manifests itself as a pronounced kink in the input-output characteristic of the nanostructure before the lasing regime is established, see for example Fig. (6) below. This kink is also associated with amplified spontaneous emission (ASE) regime, which is defined as spontaneously emitted photons amplification by stimulated emission in a single pass through the gain medium.

The lasing threshold is reached (shown by a red star) when further gain enhancement entirely compensates losses in the system, i.e., when ($R_m^{st} = D_m^{nr} + D_m^{r}$). In an experiment, this is accompanied by a phase transition from a thermal to a coherent state [216]. In the complex plane, steady-state lasing corresponds to the presence of a pole on the real axis [209], Fig. 4(b, red area). Although laser generation is essentially a nonlinear process, threshold pumping for lasing can be described through a linear approximation [64], [114], [209] because the amplitude of the laser self-oscillation is equal to zero below lasing threshold and the problem can be treated linearly, making the present description meaningful. As mentioned above, there is another, more general definition of lasing threshold, as the point where stimulated emission into the mode overcomes the spontaneous one, also known as quantum threshold condition [176], [217]. For a nanoresonator coupled to an active gain medium, this occurs when the mean photon number in the mode is unity.

The further gain increase over the lasing threshold leads to an unstable region when the laser output eventually saturates due to nonlinear effects [114], [218], Fig. 4(b, red area). The system in this regime is inherently nonlinear, this linear analysis breaks down, and stability must be carefully analyzed using nonlinear dynamics (e.g., via Maxwell–Bloch equations [200], [219], [220]). As the pole leaves the real axis moving towards the upper complex half-plane, it does not mean that lasing breaks down, because above the lasing threshold nonlinear saturation effects tend to stabilize the pole on the real axis.

If a system has no radiation loss, it does not present an amplification gap and, as a result, the lasing threshold is $R_m^{st} = D_m^{nr}$, i.e., compensation of material loss is enough for lasing. If the system has no radiation loss and material losses are small, the lasing threshold is small, which may give rise to lasers with ultralow threshold[221].

Thus, in any non-Hermitian system, the amplification regime can be achieved when net material losses are compensated, whereas the lasing threshold requires compensation of both material and radiative losses (classic definition). The two operations coincide in the absence of radiative losses, which can be achieved in different ways. First, the system can be engineered to have the vanishing radiative portion of the Purcell factor $F_r$ at the emission wavelength. In this regime, e.g., in photonic crystals with an optical band-gap or closed cavities, the system does not support spontaneous emission [179], [222], [223] and the thresholdless regime can be achieved if the system does not support other modes. Second, the system can be tailored to support so-called *bound states in the radiation continuum* (BICs) [also known as *embedded eigenstates* (EEs)], which are essentially radiationless modes of an open system [70], [224]–[226]. In the complex plane, BICs are associated with the coalescence of the zero and pole at the real frequency axis, and hence support an unbounded Q-factor and an infinitesimal linewidth in scattering experiments [64], [70], [227], [228]. If material losses are small or even absent, a small amount of gain turns the BIC into the lasing regime, which makes it promising for ultralow threshold lasers [229]–[231].

The presence of radiation loss opens a gap in the input-output characteristic (photon number-pump rate) where the amplification is possible. For example, practical optical devices like metamaterials and metasurfaces possess significant radiative damping caused by boundaries, defects, and disorder, which leads to separation of loss compensation and lasing thresholds. This gives an opportunity to explore optical amplification by gain materials, as discussed in the next section.

One more possibility enabled by active systems, which has been attracting significant attention in the nanophotonics community, is the so-called $\mathcal{PT}$-symmetry condition. A non-magnetic optical structure is $\mathcal{PT}$-symmetric if the dielectric function satisfies

$$\varepsilon(\mathbf{r}) = \varepsilon^*(-\mathbf{r}), \tag{6}$$

which imposes that the real value of permittivity is an even function of the spatial coordinates [$\varepsilon'(\mathbf{r}) = \varepsilon'(-\mathbf{r})$, while the imaginary part is odd [$\varepsilon''(\mathbf{r}) = -\varepsilon''(-\mathbf{r})$]. Interestingly, despite $\mathcal{PT}$-symmetric systems are essentially non-Hermitian, they can exhibit features similar to Hermitian systems in specific ranges of the gain/loss parameter [105], [232]. For example, the Hamiltonian of a $\mathcal{PT}$-symmetric structure can have real eigenvalues in the so-called exact $\mathcal{PT}$-symmetric phase regime. In this regime, two or more eigenvectors coincide, giving rise to an *exceptional point* (EP) [85]. On the other hand, for certain parameter ranges, the system undergoes a phase transition when the exceptional point splits into complex eigenvalues, some of them being lossy and some of them gainy so that the spectrum becomes complex, Fig. 4.

Such $\mathcal{PT}$-symmetric non-Hermitian systems provide a powerful platform for wave amplification and loss compensation in nanophotonics [86], [233]–[243]. For instance, $\mathcal{PT}$-symmetric systems can exhibit resonant localized amplification when operated in the broken phase, a phenomenon that can be exploited to obtain nonlinear responses at much lower power than in the exact phase [244] and to induce the low-threshold optical nonreciprocity [241], [242]. The $\mathcal{PT}$-symmetric structures allow single-mode lasing in systems that otherwise may support many competing resonant modes [235]. Section (VII) is dedicated to a more detailed discussion on $\mathcal{PT}$-symmetric systems.

Concluding this section, we emphasize that the classical description adopted here is applicable for linear systems, and requires caution when considering systems with large values of $\beta$. For example, in contrast to the classical definition of laser threshold (gain compensation of both radiation and dissipation losses in the mode), one should use the quantum laser threshold (number of stimulated photons $P \geq 1$, and overcomes the number of spontaneous photons in the mode). Nonetheless, the above description and classification of different possible scenarios in non-Hermitian systems with gain and/or loss provides useful insights and allows treating different effects of active nanophotonics in a single fashion.

## V. Loss compensation and amplification

Intrinsic material losses, i.e., losses that cannot be eliminated by improved fabrication techniques, hamper the metrics of nanophotonic systems for many applications. Plasmonic nanostructures provide a way to localize light below the diffraction limit [245] at significantly subwavelength dimensions, by virtue of storing part of the optical energy in the form of kinetic energy in charged carriers [246]. This makes dissipative losses inherent to any plasmonic structures [246], including highly doped semiconductors [246] and graphene [247], and the light localization in these structures comes at the price of increased dissipation. As it has been shown in Refs. [69], [248], the absorption lifetime in subwavelength metallic elements is on the order of ~10 fs in the visible range, which is much shorter than the typical absorption lifetime in dielectrics and undoped semiconductors [45]. Since dissipation losses depend on the electric field distribution in nanostructures (usually described via the confinement factor), different design approaches have been proposed [39], [249] in order to reduce the loss by tailoring the nanostructure geometry. However, this way did not demonstrate a noticeable improvement, especially in the visible range. For example, long-range surface plasmon-polaritons (LRSPPs) arising in thin metal films can have a much lower field attenuation coefficient, and hence larger propagation distance due to the symmetric mode profile in the metal, in contrast to asymmetric short-range SPPs (SRSPPs) [98].

Another approach to get rid of losses in nanophotonics is using high-index dielectrics such as Si or GaP instead of metals [41]–[53], [250]. This approach has allowed designing and implementation of low-loss nanophotonics devices, such as waveguides [251], nanoantennas [171], [250], [252], [253], oligomers [254], and has been demonstrated as a promising venue for metasurfaces [44], especially for wave-front engineering [255]. Although this approach helps to circumvent the issues of loss, such all-dielectric nanostructures obey the diffraction limit and have a moderate local field enhancement factor. Moreover, they still demonstrate non-negligible material loss, especially when coupled with quantum emitters, dropping the radiation efficiency to ~0.5 at the resonance in visible range [250].

Hence, Ohmic losses and field concentration impose fundamental challenges in plasmonic and all-dielectric nanophotonics. Due to its importance, loss compensation has been an object of active studies for a long time. Different methods for loss compensation have been proposed, which employed gain materials [72], [98], [256], [257], nonlinear effects [258], tailoring of geometry [249] and coherent optical amplification [259], [260]. In this paper, we focus on using gain materials, as the most common approach.

Optical amplification has a long history in optics, especially in the context of optical communications, where various approaches including rare-earth-doped gain materials (material gain of few cm$^{-1}$), semiconductor materials (several hundreds of cm$^{-1}$), or nonlinear effects such as stimulated Raman and Brillouin scattering, four-wave mixing, and optical parametric amplification have been successfully utilized[261]. Although active nanophotonics often borrows these approaches, here we do not focus on this well-established area. As mentioned above, if the amplification and lasing thresholds of a nanostructure do not coincide, the corresponding gap in material pumping opens a window for the net amplification over loss compensation, when the optical signal can be not only compensated but also amplified before arising lasing. For this gap to exist, the structure has to have noticeable radiation losses. A good discussion of this issue and of the possibility of loss compensation and amplification in metamaterials is presented in Refs. [262]-[263].

The first results on active loss compensation in spherical dielectric particles date back to the 70s of the last century [264]–[266]. It has been demonstrated that the extinction coefficient $Q_{ext} = Q_{scat} + Q_{abs}$, where $Q_{scat}$ and $Q_{abs}$ are radiation and absorption cross sections[267], can distinguish the regimes of undercompensation (lossy, $Q_{ext} > 0$), loss compensation ($Q_{ext} = 0$), and overcompensation or negative absorption (gain, $Q_{ext} < 0$). The regime where $Q_{ext} \to \infty$, or when the pole of the extinction coefficient lies on the real axis, corresponds to the lasing regime.

The first theoretical work on active loss compensation in metal structures was published in 1978 [268] by Plotz et al., where it was demonstrated that the reflection coefficient ($R$) from a *metal film* at its plasmonic resonance in the Kretschmann geometry can be enhanced up to the material loss compensation (i.e., $R_m^{st} = D_m^{nr}$) and further into the lasing regime ($R_m^{st} = D_m^{nr} + D_m^{r}$), when $R \to \infty$. Note that at some amount of gain, the scattering coefficient can turn to zero, $R \to 0$, when radiative losses equal to net dissipative losses ($R_m^{st} + D_m^{nr} \approx D_m^{r}$), giving rise to the *critical coupling regime* [95], [269], which corresponds to the presence of the S-matrix zero at the real frequency axis. Later, this approach has been proposed to increase the surface plasmon polariton (SPP) propagation length [270] by Sudarkin and Demkovich, who were also the first to suggest a laser-based on SPP waves. In 2004, Avrytsky analyzed theoretically the SPPs at the interface between flat and corrugated metal (Ag) and a dielectric with strong optical amplification near the energy asymptote [271]. It has been demonstrated that the material gain of $8.07 \times 10^4$ cm$^{-1}$ is required to fully compensate for the material loss around the SPP asymptote, which is the regime with the highest field confinement and hence the loss.

More recently, the experimental realization of LRSPP plasmon propagation in metallic waveguide embedded in a fluorescent polymer with net positive gain (8 cm$^{-1}$) over macroscopic distances has been reported in Ref. [123]. In another work, the gain material has (polymer layer doped with QDs) been fabricated on top of the metal surface for gain-assisted propagation of SPPs at the telecom wavelength [272]. The increase of 27% of propagation length (which corresponds to 160 cm$^{-1}$ optical gain coefficient) has been reported. Another kind of surface plasmon–polariton amplifier has been reported in Ref. [273]. In this paper, complete loss compensation (at a pump fluence of ~ 200 µJ/cm$^2$) is observed in the time domain in the visible range with the organic gain medium [4-dicyanomethylene-2-methyl-6-(p-dimethylaminostyryl)-4H-pyran] with a material modal gain of 650 cm$^{-1}$. These experimental works demonstrate the capability of active loss compensation in waveguiding systems.

The first experiments on loss compensation and plasmon resonance enhancement in solutions of metallic (silver) nanoparticles and their aggregates in rhodamine dyes (e.g., Rhodamine 590 Chloride) have been performed in Refs. [274]–[276]. In 2003, a similar system has been explored for surface plasmon laser or *spaser*, theoretically proposed by Bergman and Stockman[277]. It has been theoretically demonstrated that the energy from an active material coupled to a metal nanostructure may be nonradiatively transmitted to plasmonic oscillations and transformed into electric field energy. The idea was to tightly confine light in the form of localized plasmons into deep subwavelength dimensions overlapping with a gain material to achieve stimulated emission and plasmon oscillations amplification or lasing. Enhanced coherent plasmon oscillations, in this case, would be a source of a giant near fields (hotspots) and coherent light at the nanometer scale [149], even beyond the diffraction limit. In contrast to all-dielectric semiconductor microlasers, the quanta of the resonator in the spaser is a plasmon (boson), instead of a photon, but it can also generate coherent light emission, which has been realized in a series of subsequent experiments[39], [78], [278], [279].

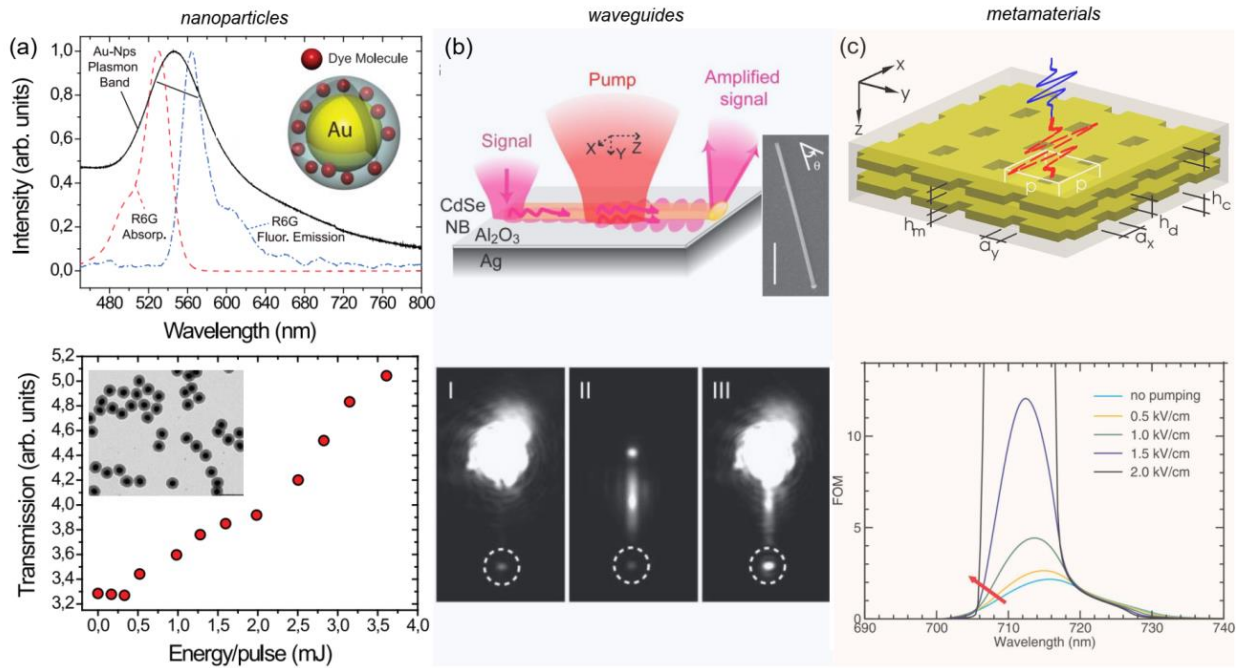

**Figure 5. Loss compensation in nanophotonic structures.** (a) Upper: Absorption (red dashed curve) and emission (blue dash-dot curve) spectra of Rhodamine 6G dye and the plasmonic resonance (black solid curve) of Au nanoparticles. Lower: Transmission enhancement vs. pumping energy[280]. Inset: SEM of core-shell nanoparticles. (b) Upper: Schematic of the hybrid plasmonic waveguide (CdSe NR/Al$_2$O$_3$/Ag) in the pump-probe setup. Inset: SEM image of the waveguide deposited on the alumina coated Ag film (the scale bar is 2 um). Lower: Optical images, corresponding to a probe signal launched from the top end of the NR waveguide and emitted from the bottom (I), PL with the pump only (II), and the amplification of the probe signal when both pump and probe are present (III)[281]. (c) Upper: Illustration of the double-fishnet structure embedded in a dielectric host material with the dye molecules. Pump (red dashed line) and probe

(blue solid line) pulses illustrate the pump-probe configuration. Lower: Calculated figure-of-merit (FOM) for different pumping amplitudes[256].

It has been soon realized that, in order to achieve an appropriate gain parameter for loss compensation in plasmonic nanostructures, the concentration of gain material has to be increased significantly in the area of most field enhancement (~ a few $10^{th}$ nm from the metal surface) [280]. At the same time, it is known that the emission of a dipole source located nearby the metal surface gets dissipated because of *quenching,* i.e., strong dissipation via non-radiative higher-order modes of a resonator [282]. At the same time, the small distance to the nanoparticle surface fulfills the requirements for effective energy transfer to the right mode to be amplified. Hence, there should be a trade-off between these two effects, which has been reported in Ref. [283] to be ~3 nm for Au/silica core-shell structure with R6G, where the time-resolved PL spectroscopy detected the minimum lifetime of R6G. This result demonstrates that loss compensation dependents critically on the interplay of the spectral overlap and the distance between plasmon and gain material [283].

In Ref. [280], suspension of densely-packed structures consisting of an Au core and $SiO_2$/R6G shell has been realized, Fig. 5(a). For a mixture of these core-shell particles in water, the enhancement of the probe beam (@532 nm) intensity as a function of pump light energy (@355 nm) has been measured, see Fig. 5(a, lower panel). The results demonstrate significant growth in both scattered intensity and transmission coefficient at the probe wavelength as the pump energy increases, Fig. 5(a, lower panel).

In addition to plasmon dipole resonances in metallic particles, the magnetic dipole Mie resonance in all-dielectric nanoparticles and nanostructures has been raising significant attention in this context, because it enables a new way of enhancing light-matter interactions, including magnetic Purcell factor [164], low-loss sensing [284], efficient second and third harmonic generation[44], [285], and wave-front engineering. It has been proposed recently that the use of gain materials can further enhance the magnetic response in dielectric nanoparticles [286].

All-dielectric and hybrid nano-waveguides represent another important class of optical nanostructures [287]–[289]. Such nano-waveguides allow light transmission in the form of tightly confined waveguiding modes and hence are of interest for optical interconnections on a chip. These structures also suffer from material and radiation loss and hence loss compensation in nano-waveguides has also been developed [123], [272], [273], [281], [290]–[302]. For example, *in situ* loss compensation has been realized in Ref. [281] in a hybrid metal-dielectric waveguide (nanorod,

NR) at room temperature, Fig. 5(b). The structure represents a CdSe long NR arranged on a metallic (Ag) substrate. It has been demonstrated that *in situ* CdSe NR has a high optical gain parameter of ~6755 cm$^{-1}$, which is enough for propagation loss compensation and waveguiding for a long distance, Fig. 5(b, lower panel). In [292] it has been also demonstrated that surface plasmon polariton amplification in a similar hybrid geometry can be realized via electrical injection.

Finally, optical *metamaterials*, or artificial electromagnetic material whose optical properties are tailored by nanostructuring, rely on light propagation through bulk optical structures that strongly interact with light and hence are very sensitive to the presence of material loss [68]. Several theoretical papers on loss compensation in metamaterials were published so far. For example, in [303] loss compensation in metamaterials composed of a pair of electric and magnetic resonators utilizing active inclusions or gain material has been discussed. Loss compensation in metamaterials composed of core-shell nanoparticles with active cores (QDs) has been also theoretically discussed in [197], [304]. Loss compensation in plasmonic metamaterials involving gain materials has been investigated in a large number of subsequent papers [94], [256], [305]–[311].

As a good example, let us consider the analysis of loss compensation by a gain material in a *negative refractive index metamaterial* in the form of two perforated silver films (double-fishnet) embedded in a dielectric host material with dye molecules [Fig. 5(c, upper panel)] reported in Refs. [256], [309], [310]. The analysis is based on a full-vectorial three-dimensional Maxwell-Bloch approach. The retrieval analysis of the effective refractive-index [$n(\lambda) = n' + in''$] extraction has been used for the estimation of the figure of merit [FOM=$|n'(\lambda)/n''(\lambda)|$] of these materials. The results presented in Fig. 5c (lower panel) show that an increase of pump amplitude from 0.5 kV/cm to 1.5 kV/cm increases the FOM from a value of 3 (@716 nm) to 12 (@712 nm).

Further increase in pumping leads to further increase in FOM. Loss compensation in a negative-index fishnet metamaterial has been realized in Ref. [72]. In this work, an optically pumped gain material (Rh800) combined to a plasmonic metamaterial in the range between 722 and 738nm has been used to demonstrate loss compensation. In this range, the metamaterials become active, as the sum of the light intensity measured in transmission and reflection exceeds the intensity of the incident beam. The response of a larger number of layers has also been theoretically analyzed [312]. Next, in Refs. [313], [314] loss compensation using optically pumped semiconductor QDs coupled to the plasmonic metamaterials has been experimentally demonstrated. In these works, the gain effect is manifested as a multifold PL intensity increase and spectral narrowing. Loss compensation in negative-index metamaterials via optical parametric amplification (OPA) has been suggested in [258].

*Hyperbolic metamaterials* are another interesting class of metamaterials. Because of their hyperbolic (or indefinite) dispersion, they possess (theoretically) unbounded values of the local density of states and hence ultra-strong, and broadband, light-matter interactions, like Purcell effect [315]. Since these anisotropic materials require dielectric permittivity with different signs of the real part, they intrinsically require using inherently lossy metals or doped semiconductors. Loss compensation in hyperbolic metamaterials has been studied in Refs. [196], [260], [316], [317](see also review paper [318]). Although theoretical studies demonstrate the ability of loss compensation in hyperbolic metamaterials along with significant improvement of their performance (for example, producing narrower focal points than their passive counterparts), there are no reported convincing experimental results to date.

Thus, over the nearly 40-year history of research in the field of loss compensation in dielectric and plasmon nanostructures, tremendous work has been done. It was shown that the material gain required for loss compensation of LRSPPs in thin metal films is $\sim 10^2 - 10^3$ cm$^{-1}$, of SISPPs is $2 \cdot 10^3 - 5 \cdot 10^3$ cm$^{-1}$, and the highest required gain is for SPPs near the energy asymptote (up to $10^5$ cm$^{-1}$). Loss compensation in plasmonic nanoparticles requires a material gain of $10^3$ cm$^{-1}$. Dielectric structures, nanoparticles, and waveguides require significantly smaller material gain due to their lower material loss ($\sim 10$ cm$^{-1}$ for Si wavegiudes and $\sim 10^2$ for high-index nanoparticles). Moreover, semiconductor nanostructures often allow the realization of the gain *in situ*. These gain levels are typical for traditional gain materials (see Figure 2) in both optical and electrical pumping.

## VI. Lasing

Since the invention of lasers [319], there has been continuous progress in their miniaturization, lasing threshold reduction, and ultrafast operation. In this direction, various kinds of dielectric lasers have been realized, including vertical-cavity surface-emitting lasers [320], microdisk lasers [321], Fabry-Perot lasers [322], photonic crystal lasers, and others. Although these systems can offer low lasing thresholds, their dimensions are limited to several operating wavelengths. It is generally believed that shrinking the laser size down will lead to a significant decrease in the laser threshold, high modulation speeds, and single-mode operation[74]. In this section, we discuss the state-of-the-art optical nanolasers and summarize the successes and trends in this area.

The first theoretical works on light scattering from spherical dielectric nanoparticles with gain and observation of singularity-like scattering response date back to 1978 [265], [266]. However, the connection between these singularities and lasing has been established relatively recently [323]–[325]. It was realized that the size of dielectric resonators is restricted by the diffraction limit [$\sim(\lambda/2n)^3$, $\lambda$ is the resonant wavelength, $n$ is the resonator refractive index] that does not allow to make an essentially subwavelength laser.

As plasmonics has become mainstream, it has been suggested that the size of lasers can be reduced to subwavelength dimensions with the aid of surface plasmon-polariton modes [39], [67], [74], [75], [78], [94], [191], [256], [277], [278], [326]–[328]. Lasing in such plasmonic-based structures relies on population inversion of emitters (e.g., fluorophores, QDs) and feedback provided by plasmonic resonant modes. The concept of a spaser [75], [78], [79], [81], [82], [192], [329], amplifier of localized surface plasmons oscillating in metal nanoparticles arranged in a dielectric environment, further generalized to include traveling surface plasmon polaritons, turned out to be of great interest for nanophotonic applications. In addition to its vital importance for advanced optoelectronics, today, they have been applied for sensing [76], [330] and biological imaging with high resolution [74]. The subsequent invention of new kinds of nanoresonators and active materials led to a family of nanolasers attracting great attention from the nanophotonics community [39], [78], [192], [277]–[279], [331]–[339].

Since the invention of plasmonic nanolasers, different approaches and geometries have been explored, including metallic-nanoparticle lasers [278], [340], plasmonic nanowire lasers [67], [78], coaxial nanolasers [180], metal-insulator-metal gap-mode nanolasers[335], plasmonic crystal nanolasers [341], Tamm plasmon laser [185] to name just a few. We address the interested readers to the review papers discussing the various nanolaser geometries [74], [342]. Below in this section, we discuss the problem of lasing threshold definition, different laser geometries, and new active materials (2D TMDCs, perovskites, MOFs), perspective for nanolaser realizations.

*Lasing threshold.* It has been generally assumed for a long time the lasing regime can be reached when the right amount of gain compensates both net radiation and material losses in a system (the classic definition of lasing threshold). In the complex frequency plane, lasing is associated with the appearance of a pole of the S-matrix (or scattering/reflection amplitude), which moves from the lower complex half-plane towards the real frequency axis, as the gain increases [64]. The distance of the pole (its imaginary frequency) from the real axis depends on the total loss in the system and defines the decay rate (and *Q*-factor) of the mode. Thus, lasing requires the presence of a mode (pole) and an active medium that compensates the radiative and material losses, which is realized via feedback for a sufficiently large confinement factor, $\Gamma$.

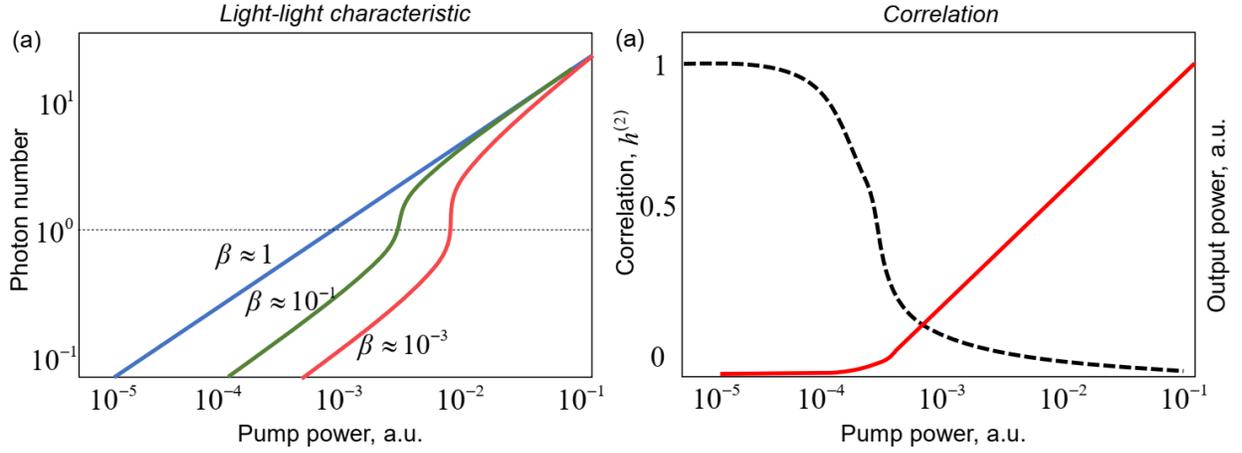

**Figure 6.** (a) Typical dependence of light-light characteristic or number of protons in the mode ( $P$ ) as a function of pump power for different values of $\beta$-factor. (b) Typical dependence of second-order autocorrelation function [ $h^{(2)} = g^{(2)}(0) - 1$ ] (black dashed curve) and output power (red solid curve) as a function of the pump power. The laser regime manifests itself as a nonlinear growth of output power (kink) in (a) and reducing the correlation function $g^{(2)}(0)$ from thermal state [ $g^{(2)}(0) = 2$ ] to coherent state [ $g^{(2)}(0) = 1$ ]. In general, the nonlinear kink and the correlation function are independent, meaning that the kink is not associated with the right laser statistics.

The classic lasing threshold is defined as the point in which the total dissipation rate ( $D_m^{nr} + D_m^r$ ) of the resonator mode equals the gain ( $R_m^{st}$ ), i.e., the stimulated emission rate to the mode (see Section III). This lasing threshold can be reduced by decreasing losses, hence increasing the Q-factor of the resonance, which leads to smaller pumping intensities. This strategy has been realized in microlasers (size of several $\lambda$ ) based on microdisk resonators [321], [343]–[346], microspheres [347], Bragg reflector microcavities [348], and photonic crystals [349]–[351], where extremely high Q factors ($\sim 10^3 - 10^6$) can be realized. However, such relatively big lasers usually have a small $\beta$-factor ($\sim 10^{-2} - 10^{-3}$) and, as a result, a pronounced input-output characteristic (photon number-pumping rate) or "kink", see Fig. 6(a). Qualitatively, the laser threshold can be defined as a regime in which the number of stimulated photons in the mode reaches 1 ($P_{th} = 1$) as this point corresponds to the most strong nonlinearity in the input-output characteristic [dashed line in Fig. 6(a)]. This condition is also known as the *quantum threshold condition* [176], [217].

However, it has been demonstrated that this classic definition works well only for small $\beta$-factors, i.e. when the spontaneous fraction of emission in the mode is negligible. For a large $\beta$-factor, the spontaneous emission to the mode is very high, which spoils the statistics of the radiation with simultaneous disappearance of the nonlinear kink in the light-light (input-output) characteristics, see Figure 6(a) [184], [217], [352], [353].

The key property of lasing action to be investigated is the Poisson statistics of emitted photons. This statistics can be defined by the second-order correlation function, which is measured via the Hanbury Brown and Twiss experiment with a 50/50 beam splitter and two single-photon counting detectors [354]–[356]. In the stationary state, the second-order correlation function is

$$g^{(2)}(\tau) = \frac{\langle I(t) \cdot I(t+\tau) \rangle_t}{\langle I(t) \rangle_t^2}, \tag{7}$$

where $I(t)$ is the light intensity, $\tau$ is the time shift, and $\langle I(t) \rangle_t$ denotes the expectation value of the intensity in the cavity mode at the time $t$. Also, the normalized second-order correlation function can be defined as $h^{(2)} = g^{(2)}(0) - 1$. In lasers, $g^{(2)}(0)$ reduces from 2 ($h^{(2)} = 1$) for the thermal state (bunched photons) to 1 ($h^{(2)} = 0$) for the coherent state (random photons) as the pump rate increases [216], [353], [357], see Fig. 6(b). The connection between the second-order correlation function and coherence was first discussed in Ref. [358]. Note that in general, the nonlinear kink and the correlation function are independent, meaning that the kink is not necessarily associated with the right laser statistics. In lasers, $g^{(2)}(0)$ can be even less than 1 giving rise to the so-called antibunched nonclassical emission [353]. Since the second-order correlation function is rarely reported for nanolasers and the laser regime is usually determined via the apparent laser conditions, we do not discuss the question of whether the laser regime is real or only apparent in the works discussed below.

The lack of understanding of the lasing threshold still sometimes leads to an incorrect interpretation of experimental results [217]. For example, the amplified spontaneous emission (ASE) regime, defined as spontaneously emitted photons amplified in a single pass through a gain medium, is often mixed up with laser emission. This regime demonstrates many of the properties peculiar to conventional lasers, including linewidth narrowing, a nonlinear kink in the L–L curve [133], [359], and specific polarization features. Moreover, various active optical systems can demonstrate a laser-like behavior (kink, linewidth narrowing) in their pulsed regime without being a laser. These properties are called now *apparent laser conditions* [217], [360] and are not sufficient to ensure the lasing regime [217], [361]. For example, the comprehensive and critical analysis of lasing parameters of 2D TMDC-based lasers reported in [217] shows that none of these devices reported clear coherence properties and did not reach the quantum lasing threshold. A large number of erroneous interpretations of lasing in publications forced journals of the *Nature* family to introduce a list of requirements to any publication claiming new laser designs, which have to be satisfied before peer review [362].

*Nanolaser realizations*. The first attempt to realize a spaser was presented in Ref. [278]. In this work, Au nanospheres have been surrounded by a dielectric gain shell with active dye molecules [a liquid solution of Au/silica/dye core-shell nanoparticles (size of ~15 nm) of a high concentration], Fig. 7(a, upper panel). The growth of stimulated emission intensity and narrowing of the emission line with increasing pump intensity has been reported, Fig. 7(a, lower panel). The results are shown to be independent of the particle concentration indicating that the lasing coming from individual particles rather than from a collective response. However, significant material losses of such nanolasers comprising metal nanoparticles imply that the active material had a significant gain factor. For example, a recent rigorous analysis [135], [276] demonstrates that in this spaser configuration, the gain factor required to achieve the lasing threshold is unachievable in practice. Also, small metallic particles are subject to nonlocality and spatial dispersion, which can additionally increase the lasing threshold [363]. To achieve the lasing regime in metallic nanoparticles embedded in a gain dielectric environment, a material gain of ~1500–2500 $cm^{-1}$ is required [276], [364]. Thus, the lack of experiments on the second-order correlation coefficient does not allow to say with confidence in which regime this nanolaser operates.

When plasmonic nanoparticles are larger, resonant higher-order modes can appear in the radiation spectrum, and they can also lase. However, a rigorous theoretical analysis shows that the dipolar mode exhibits the lowest lasing threshold [365]. Various approaches to reduce the laser-threshold in such plasmonic nanoparticle-based spasers have been proposed, including inverted structures with dielectric gain as a core surrounded plasmonic shell [366], increase of the background permittivity [367], the use of low-loss materials (e.g., Ag instead of Au) [136], exotic shapes [77], [190], [368] and electric pumping [326], [369].

Surface plasmon polariton spasers have been demonstrated in 1D and 2D configurations [78], [100], [279], [335]. Such hybrid nanolasers have attracted significant attention because they combine the benefits of small mode volumes of plasmon modes and strong gain parameter in high-index semiconductors [370]. For example, in Ref. [78] a 1D plasmonic laser consisting of a CdS semiconductor nanowire (NW) on top of an Ag substrate has been realized, Fig. 7(b, upper panel). The gain in this structure stems from the CdS semiconductor, and its geometry supports both photonic and plasmonic modes. The photonic modes have a cut-off at $d \approx 140$ nm, while plasmonic modes exist for any $d$ and are localized between the high-index NW and the silver surface, with an ultra-small mode area ($\sim \lambda^2 / 400$), resulting in strong electric field enhancement in the lasing mode. The four spectra for different peak pump intensities reported in Fig. 7(b, lower panel) exemplify the transition from spontaneous emission (21.25 MW cm$^{-2}$) via amplified spontaneous emission (32.50 MWcm$^{-2}$) to full laser oscillation (76.25 MWcm$^{-2}$ and 131.25 MWcm$^{-2}$) with the observation of a typical kink in the input-output characteristic, Fig. 7(b, inset in the lower panel). The geometrical parameters are chosen to be d=129 nm, h=5 nm, so that the structure has only plasmonic and no photonic modes. Interestingly, the value of $\beta$-factor in this structure reaches 0.8 for a small gap between NW and Ag substrate, which manifests itself in a low lasing threshold (Fig. 7b, inset in the lower panel).

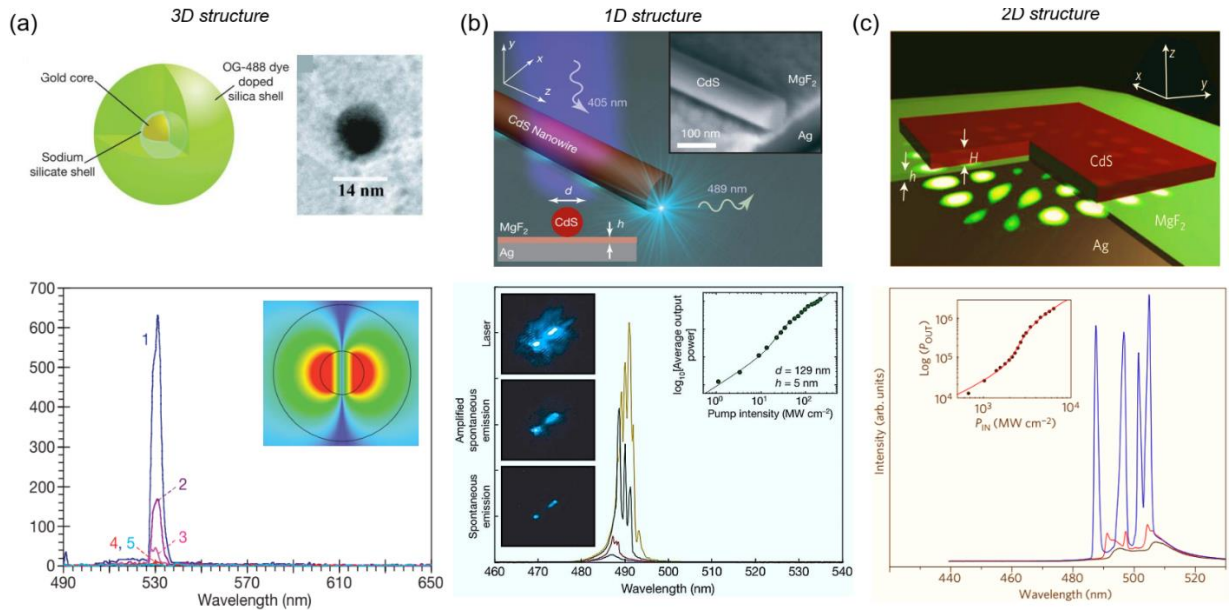

**Figure 7**. **3D, 1D, and 2D spasers**. (a) Upper: 3D spaser with dye molecules incorporated into the silica shell. Inset: SEM image of the realized spaser. Lower: emission spectra of the sample pumped with 22.5 mJ (1), 9 mJ (2), 4.5 mJ (3), 2 mJ (4) and 1.25 mJ (5) 5-ns laser pulses (@488 nm). Inset: electric field distribution of the laser mode[278]; (b) Upper: 1D plasmonic laser consists of a CdS semiconductor NW on top of an Ag substrate, separated by a nanometer-scale MgF$_2$ layer of thickness h. Inset: SEM image of the fabricated structure. Lower: Four spectra for different peak pump intensities exemplify the transition from spontaneous emission (21.25 MW cm$^{-2}$) via amplified spontaneous emission (32.50 MWcm$^{-2}$) to full laser oscillation (76.25 MWcm$^{-2}$ and 131.25 MWcm$^{-2}$), for d=129 nm, h=5 nm (longitudinal modes). Inset: Dependence of output power on the peak pump intensity[78]. (c) Upper: 2D plasmon laser consisting of a thin CdS square atop a silver substrate separated by a 5 nm MgF$_2$ gap. Lower: Room-temperature laser spectra and integrated light-pump response (inset) showing the transition from spontaneous emission (1960 MWcm$^{-2}$, black) through amplified spontaneous emission (2300 MWcm$^{-2}$, red) to full laser oscillation (3074 MWcm$^{-2}$, blue). Inset: The nonlinear response of the output power to the peak pump intensity [279].

A 2D SPP laser with subwavelength vertical dimensions has been realized in Ref. [279]. This laser consists of a thin CdS square (45-nm-thick) atop an Ag substrate, separated by a 5 nm MgF$_2$ gap, Fig. 7(c, upper panel). Like in the previous work, the gain in this structure stems from the CdS semiconductor. Room-temperature laser spectra and integrated light-pump response (inset) showing the transition from spontaneous emission (1,960 MWcm$^{-2}$, black) through amplified spontaneous emission (2,300 MWcm$^{-2}$, red) to full laser oscillation (3,074 MWcm$^{-2}$, blue) have been reported, Fig. 7(c, lower panel), with a characteristic kink in the input-output characteristic (inset). Interestingly, by controlling the shape of the CdS particle, the structure is tailored to support single-mode lasing at the wavelength of 495 nm. A similar structure (CdS square on top of an Au substrate) has been comprehensively studied in Ref.[175] and compared to the same but placed on SiO$_2$ substrate (photonic nanolaser). It has been experimentally demonstrated that the Au nanolasers have unusual scaling laws allowing them to be more compact and faster with lower power consumption when their cavity size approaches or surpasses the diffraction limit, whereas the photonic nanolasers do not demonstrate such a performance.

In addition to their nanoscale dimensions, nanolasers have been demonstrated to have an *ultrafast response*, caused by the accelerated spontaneous emission (Purcell effect) into the lasing mode, which allows emission of ~1 ps pulses of coherent light. The Purcell effect also increases the $\beta$ factor (a portion of energy going to the lasing mode, see Section II), which further accelerates the transient response and reduces laser-threshold[74]. Further developments in the area of nanolasers are associated with searching for *new semiconductor materials* with higher material gain and *fundamentally new lasing modes*. Below we discuss these current trends in nanolasers research.

*New gain materials.* The first class of prospective materials for nanolasers considered here is single-layer transition metal dichalcogenides (TMDCs). These 2D materials (MX$_2$, M = Mo, W; X = S, Se) are semiconductors and, in contrast to the graphene (semimetal), have energy bandgaps spanning the visible and near-infrared spectral region (1.0–2.1 eV). 2D TMDCs in the monolayer limit demonstrated a strong PL, attributed to the transition from indirect to direct bandgap emission [371], [372]. Due to weak dielectric screening, excitons in 2D TMDCs have large bonding energy, about one order of magnitude larger than the thermal energy at room temperature [373]. These fascinating properties make 2D TMDCs of interest for various photonics and optoelectronic applications [372]–[383].

Nanolasers based on TMDC materials have been realized recently in Refs. [384]–[387]. For example, in [384] a structure composed of a monolayer WSe$_2$ material acting as a gain medium placed on a photonic crystal cavity was reported, Fig. 8(a, upper panel). The lower panel shows the dependence of output intensity at 740 nm on the optical pump power at 130 K, cavity emission (red filled squares), spontaneous emission off cavity resonance (violet half-filled squares), simulated curves using the laser rate equation (solid lines). The dark grey dashed line corresponds to the defined lasing threshold. The laser operation regime is observed with an optical pumping threshold as low as 27 nW or ~1 W cm$^{-2}$. A similar optical pump threshold has been reported in [388] (~1.3 Wcm$^{-2}$) for a single layer of QDs photonic crystal cavity laser. Such a low lasing threshold is a consequence of the 2D nature of the monolayer WSe$_2$.

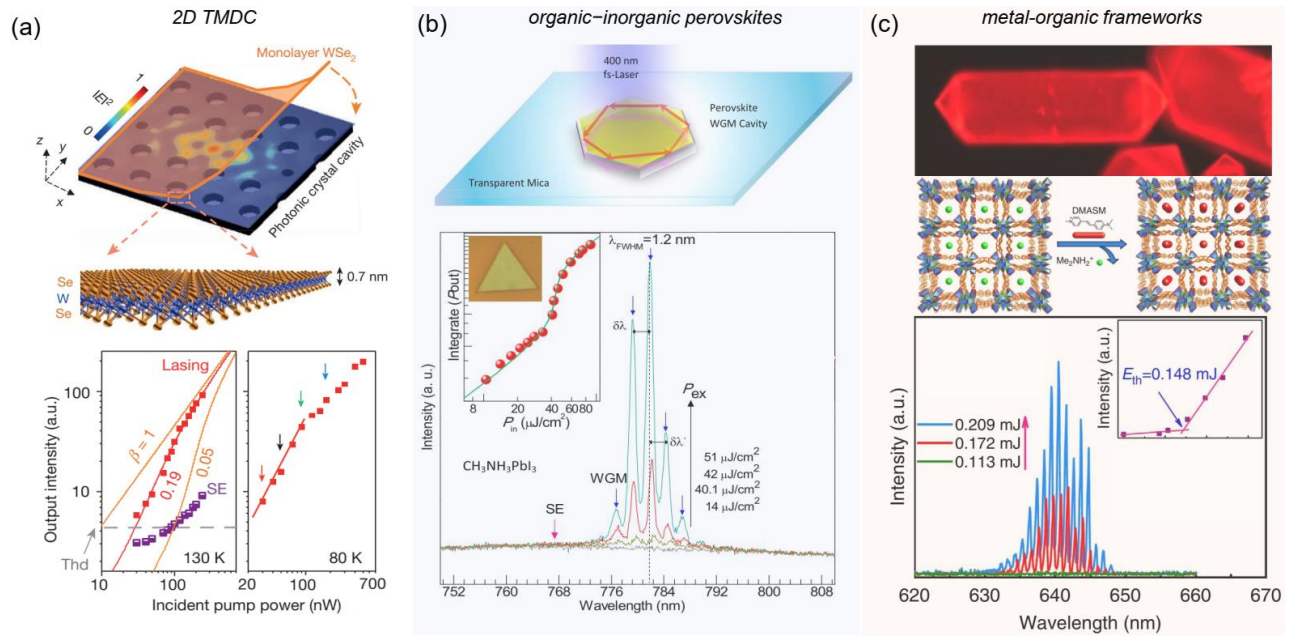

**Figure 8. New materials for nanolasers**. (a) Upper: Hybrid monolayer WSe$_2$– photonic crystal cavity nanolaser architecture. Inset, schematic of the atomic structure of monolayer WSe$_2$. Lower: (left) dependence of the light output intensity (@740 nm) on the optical pump power at 130 K, cavity emission (red filled squares), spontaneous emission off cavity resonance (violet half-filled squares), simulated curves using the laser rate equation (solid lines), dark grey dashed line corresponds to the defined laser threshold. (right) input-output characteristics for the same device at 80K (red squares)[384]. (b) Upper: Schematic of the perovskite-based laser. Lower: Evolution from spontaneous emission (SE, ~768 nm) to lasing (whispering-gallery modes indicated by blue arrows) in the CH$_3$NH$_3$PbI$_3$ triangular nanoplatelet (thickness is 150 nm; edge length is 32 μm). For pumping, the pulsed laser (400 nm, 150 fs, 1 kHz) is used. The pumping fluence (P$_{ex}$) from low to high is 14 (gray), 40.6 (dark yellow), 42 (red), and 51 (olive) μJ/cm$^2$, respectively. Inset:

integrated output emission (Pout) over the whole spectral range as a function of $P_{ex}$ [389]. (c) Upper: Fluorescence microscopic images and emission spectra of bio-MOF-1&DMASM with dye content of 40.0% illuminated with ultraviolet light. Middle: Schematic of the process of encapsulation of pyridinium hemicyanine cationic dye DMASM into bio-MOF-1. Lower: two-photon-pumped lasing spectra of bio-MOF-1&DMASM under different pumped pulse energy. Inset: input-output characteristic of the laser [390].

Existing semiconductors with high quantum efficiencies, like CdS discussed above, possess enough material gain for lasing at room temperature in the visible range. In turn, 2D TMDC-based lasers are complicated in terms of fabrication because they require high Q-factor resonators and precise alignment. Lasers based on so-called *organic-inorganic perovskites*, which are another interesting new material for nanophotonic lasers, allow overcoming these difficulties. These materials have substantial material gain, high quantum efficiency, and high-index permittivity, which make it possible to realize the *in-situ* lasing scenario. Importantly, these organic-inorganic perovskite materials can be fabricated with cost-effective wet chemistry methods at room temperature, which makes fabrication straightforward. Fig. 8(b) demonstrates an example of perovskite ($CH_3NH_3PbI_3$)-based laser. This material is attributed to a more general class of methylammonium lead halide perovskites $CH_3NH_3PbI_3$-aXa (X = I, Br, Cl) ($0 \leq a \leq 3$), demonstrating optical pumping lasing threshold ($P_{th}$) of ~ 37 $\mu J/cm^2$, Fig. 8(b, lower panel). Interestingly, a simple chemical changing of X or *a* leads to the tuning of the PL emission spectrum and, as a result, a lasing wavelength spanning a wide range of frequencies in the IR spectrum. Due to their tunability, relatively low lasing threshold and *in-situ* lasing scenario, organic-inorganic perovskite nanolasers have raised significant attention recently [140], [389], [391]–[398].

The last perspective gain material for laser purposes that we discuss here is *metal-organic frameworks* (MOFs), also known as porous coordination polymers, which are 3D ordered porous materials composed of metal ions/clusters connected by organic ligands (linkers). These materials have substantial internal surface areas, providing a space for encapsulation of guest molecules/ions, which can significantly change and enrich their optical properties. The luminescent properties of MOFs can stem from organic ligands and guest molecules. For example, the encapsulation of organic dyes into the pore spaces of MOFs leads to immobilization of them and minimizes the aggregation-caused quenching. The rapid progress in this area during the past few years led to fascinating applications of MOF materials and nanocrystals in light-emitting devices [399].

Recently, these materials have been utilized for advanced lasers. Fig. 8(c) demonstrates one of these recent realizations [390]. In this work, the cationic pyridinium hemicyanine dye DMASM encapsulated in the pores of MOF with large two-photon absorption is used as a gain material, whereas the relatively high refractive index of the MOF matrix is used as a laser cavity with Q factor ~1500 [Fig. 8(c, upper panel)]. The structure demonstrates lasing around 640 nm when pumped under a 1064-nm pulse laser with a threshold of ~0.148 mJ, Fig. 8(c, lower panel).

Concluding this discussion, we stress that the materials discussed here do not necessarily cover the whole range of new advanced materials for nanolasers. For example, a low-lasing threshold in so-called *organic semiconductor*-based lasers has also been recently reported (see, e.g., [400]–[402]).

*New lasing approaches*. There are different approaches to low-threshold lasers. One approach relies on group velocity engineering when in the slow-wave regime radiation leakage from the cavity significantly drops [403]. The group velocity vanishes around *van Hove singularity* [404], [405], resulting in a large local density of states and Purcell factor. This results in strong light-matter interactions, low radiation losses, and hence a low lasing threshold. A combination of this approach with a single-mode response turns the $\beta$ - factor to large values (up to 0.97 and higher) [181].

In order to reduce radiation losses and hence reduce the lasing threshold, one can engineer the structure supporting resonant dark modes [336], [406]. An excellent example of such high Q-factor states is represented by the strong collective resonances in 2D plasmonics arrays [407]–[409]. An ultimate approach in this context consists in utilizing fundamentally uncoupled states from the continuum of radiation modes or BICs [62], [64], [70], [225], [226]. In the ideal case, BICs have no radiation losses, and hence the Q-factor of a structure without no material loss can be unlimitedly large [64]. Therefore, such structures are promising for low-threshold lasers [62], [230]. For example, in [410] the lasing regime (at low temperature, 78 K) in a periodic structure composed of coupled gallium arsenide (GaAs) nanopillars supporting symmetrically-protected BIC at normal incidence ($\Gamma$ point of the first Brillouin zone) has been reported. The device emits light in the laser regime at ~825 nm being pumped by a femtosecond laser (780 nm, 200 fs pulse width at a repetition rate of 100 kHz) with the measured lasing threshold of 20 μJcm$^{-2}$. Even though GaAs has a relatively low material gain, the Mie resonances in the GaAs nanoparticles boost the gain to sufficiently large values for demonstration of stimulated emission.

Lastly, topologically nontrivial lasers have been attracting a great deal of attention since their recent discovery [411], [412]. Photonic topological insulators are a novel phase of artificial materials that support unidirectional protected edge states arising in the photonic bandgaps of tailored periodic structures. These states are robust to disorder and fabrication imperfections and hence provide an unprecedented platform for robust photonic devices [107], [413]–[417]. In [411], [412], it has been demonstrated that an array of wisely designed micro-resonators supporting photonics topological edge state provides not only robust behavior of the lasing edge mode with respect to defects but also can possess a considerably higher slope efficiency as compared to ordinary nonprotected modes. The most recent works on topologically enabled ultra-high-$Q$ resonances robust to imperfection and hence ideal for laser applications [418]. These works open new directions in the field of active nanophotonics.

## VII. $\mathcal{PT}$-symmetry and exceptional points

As mentioned in Section (II), a special class of active nanophotonic structures is $\mathcal{PT}$-symmetric structures. The $\mathcal{PT}$ symmetry concept dates back to the pioneering works of Bender and Boettcher in theoretical quantum physics [88], [419]. It has been shown that a large class of non-Hermitian Hamiltonians can exhibit entirely real spectra, as long as they respect $\mathcal{PT}$-symmetry, i.e., symmetry in the inversion of time ($\mathcal{T}$) and space ($\mathcal{P}$) [88], [89], [232], [419]. Recently the concept of $\mathcal{PT}$-symmetry was extended to optical systems [87], [420], [421], where time-reversal (complex conjugation) involves mapping optical gain into loss[90].

Arguably the simplest example of a $\mathcal{PT}$-symmetric structure is a coupler composed of two evanescently coupled guided-wave elements, i.e., optical waveguides or cavities, which are identical in any sense except for the presence of gain in one element while the other element exhibits the same amount of loss, see Fig. 1(e). The elements are coupled to each other with coupling strength $\kappa$. If the coupling is weak, the system possesses two modes, one lossy and one amplifying. When excited, the lossy mode decays exponentially in time, whereas the gainy one exhibits exponential growth. However, if the coupling strength is large enough, the system resides in the strong coupling regime when the coherent energy exchange between the elements compensates for the loss decay stabilizing the system at the exceptional point (EP) at the real frequency axis. Thus, the EP in active non-Hermitian structures is characterized by coalescence of two (or more) poles at the real frequency axis [64].

Thus, the behavior of $\mathcal{PT}$-symmetric structures is governed by balanced gain/loss, or Hermiticity, parameter ($g$). For small $g$, the coupling rate between individual components is sufficiently strong compared to the gain and loss rates ($\mathcal{PT}$-symmetrical phase), leading to power compensation and $\mathcal{PT}$-symmetric eigenmodes have real eigenfrequencies [88]. For sufficiently large $g$, the systems undergo a phase transition at an EP. Past this point, the eigenvalues become complex, moving to the regime of spontaneously broken $\mathcal{PT}$-symmetry [422]. In this scenario, the eigenstates are no longer invariant under simultaneous parity and time-reversal but turn one into another in spite of the fact that the governing Hamiltonian commutes with the $\mathcal{PT}$ operator [87], [88], [419], [423]. At the exceptional point singularity, not only the eigenvalues but also their associated eigenvectors, coalesce. If $N$ is the number of coalesced modes, then the EP is of $N^{\text{th}}$ order. One of the most fascinating and promising features that make EPs very attractive from the application viewpoint is that an EP-based sensor with order $N$ possesses the sensitivity of $\sim \varepsilon^{1/N}$, with $\varepsilon$ being a small parameter (variation). As a result, EP-based sensors may exhibit higher sensitivity compared to a conventional sensor with sensitivity $\sim \varepsilon$ [424]–[429]. We note, however, that the inherent quantum noise of active systems implies a limitation of sensitivity of such sensors, which opens a debate on whether this approach truly enhances sensitivity in practical systems [424], [430]–[432].

Beyond these peculiar spectral properties, $\mathcal{PT}$-symmetric systems can also exhibit unique scattering properties, such as unidirectional or bidirectional invisibility, or concurrent coherent perfect absorption (CPA) and lasing [64], [237], [238], [433]–[437]. $\mathcal{PT}$-symmetric structures exhibit strongly asymmetric reflection when operating near the exceptional point [92], [438]. This behavior can be exploited for asymmetric interferometric light-light switching [439] and one-sided perfect absorption [440]. In [439], a $\mathcal{PT}$-symmetric metawaveguide was realized to operate close to the exceptional point, where strong reflection asymmetry allowed for a CPA eigenmode with strongly asymmetric incidence. In turn, this property enabled a strong signal beam that could be manipulated by a much weaker control beam. An extinction ratio of up to 60 dB was observed, with a 3:1 intensity ratio between signal and control beam.

An interesting example of one-dimensional $\mathcal{PT}$-symmetric structure is a periodic arrangement of gain and loss slabs, which exhibits interesting properties, such as *unidirectional invisibility* [64], [92], [438], [441]. Modes in the unbroken $\mathcal{PT}$ symmetry regime live equally in the gain and loss regions and thus remain neutral, while for a pair in the broken $\mathcal{PT}$ phase regime, one mode is located mostly in the active cavity while the other mode remains confined to the lossy cavity [442]. As a result, through proper design of the phase transition threshold in the system, one of the broken-symmetry modes can lase while all other modes being prevented from lasing [86], [443]. The mode selection property of parity-time symmetric systems has been experimentally demonstrated in optically-pumped semiconductor coupled microring lasers [86], opening interesting opportunities for active nanophotonic systems. In such a scenario, gain and loss are controlled through selective pumping of the microring resonators. In a different effort, a microring laser with periodic modulation of loss segments has been also demonstrated [443].

## VIII. Conclusions and Outlook

In this paper, we have reviewed the current stage of research and recent efforts in the broad field of active nanophotonics. We have discussed available approaches and materials for active nanophotonics, provided a theoretical introduction on the topic, and analyzed different possible scenarios of non-Hermitian systems, including loss compensation and amplification, lasing, stability and $\mathcal{PT}$-symmetry. The bulk part of this work is dedicated to the state-of-the-art of loss compensation and lasing in nanostructures of different dimensions and materials. We have also covered recent studies on advanced materials (2D TMDCs, perovskites) for active nanophotonics and novel concepts, including bound states in the continuum, $\mathcal{PT}$-symmetry, exceptional points, and nontrivial lasing. Yet, this review is by no means exhaustive and we left untouched several important research directions of active nanophotonics. First, we did not discuss the laser operation in the so-called strong-coupling regime (with a gain-resonator coupling coefficient larger than the total decay rate), when excitons in gain media and polaritons in the cavity mode cannot be distinguished any longer, and exchange energy coherently via Rabi oscillations. In this regime a new kind of quasiparticle, exciton-polariton, emerges causing different operation mechanisms compared to conventional lasers [444].

Second, we briefly discussed the emerging topic of topologically nontrivial lasers [411]. Topological insulators with nontrivial topological properties of their photonic band structure have become a subject of intensive research efforts in nanophotonics. These topologically protected surface states exhibit also the remarkable property of spin-locked propagation with forbidden back-scatterings, which is becoming of great interest for electronics, optoelectronics, photonics, spintronics, and quantum computation.

Another attractive state of matter with unusual topological properties are Weyl semimetals (WS), which have raised significant attention due to their massless bulk fermions and topologically protected Fermi arc surface states [445]. Weyl semimetals' band structure embraces an even number of nondegenerate band-touching points arising at the Fermi level (Weyl nodes) split in the momentum space, giving rise to their topological nature. The Weyl points appear in pairs, exhibit a linear dispersion around them, and can have positive or negative helicity charges. The existence of the defined helicity charge unites Weyl points with helicity degenerate valleys in 2D transition metal dichalcogenides (TMDC) recently emerged as a versatile platform for valleytronic applications [383]. Recently, similar concepts to topological insulators and Weyl semimetals have been realized in photonics, giving rise to topologically nontrivial photonics [413], [414]. The studies on active properties and lasing in these structures have recently raised strong interest, demonstrating promising results on single-mode lasing robust against defects [411], [412]. We envision that the combination of these concepts, and the inspiration from recent advances in quantum mechanics and condensed matter systems, will continue to drive the evolution of the field of active nanophotonics, leading to remarkable technological advances for low-energy, ultrafast optoelectronic and photonic systems.

## Acknowledgments

We thank Dr. Denis Baranov (Chalmers University of Technology) and Dr. Andrey A. Vyshnevyy (Moskov Institute of Physics and Technology) for useful discussions. We acknowledge support from the Air Force Office of Scientific Research, the Office of Naval Research, the National Science Foundation, and the Simons Foundation.

## About the Authors

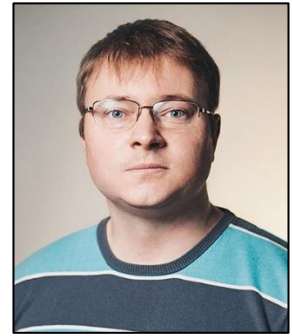

**Alex Krasnok** (Member, IEEE) earned his Ph.D. with distinction from ITMO University (St Petersburg, Russia) in 2013. After spending two years (2016-2018) as a research scientist at The University of Texas at Austin (Austin, USA), he joined CUNY Advanced Science Research Center (New York, USA) as a Research Assistant Professor and Core Facility Director. His current research interests are in the areas of applied electromagnetics, nanophotonics, metamaterials, plasmonics, and nanotechnology, with particular emphasis on cross-disciplinary research. He has made significant contributions in the areas of extreme scattering engineering, nanoantennas, metasurfaces, optics of 2D transition-metal dichalcogenides, and low-loss dielectric nanostructures. He has authored or co-authored five books and book chapters, four patents, and more than 120 papers. He has earned several research awards and grants, including the gold medal of Nobel Laureate Zhores Alferov's Foundation (2016).

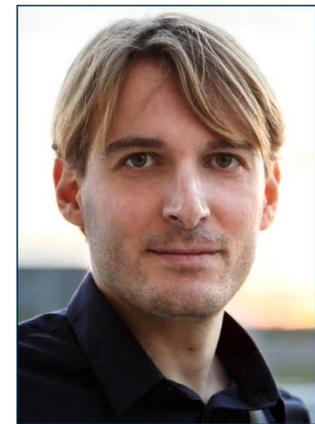

**Andrea Alù** (Fellow, IEEE) received the Laurea, M.S., and Ph.D. degrees from the University of Roma Tre, Rome, Italy, in 2001, 2003, and 2007, respectively.

Andrea Alù is the founding director of the Photonics Initiative at the CUNY Advanced Science Research Center, Einstein Professor of Physics at the CUNY Graduate Center, and Professor of Electrical Engineering at The City College of New York. He is affiliated with the Wireless Networking and Communications Group and the Applied Research Laboratories, both based at the University of Texas at Austin, where he is a Senior Research Scientist and Adjunct Professor. His research interests span over a broad range of technical areas, including applied electromagnetics, nano-optics and nanophotonics, microwave, THz, infrared, optical and acoustic metamaterials and metasurfaces, plasmonics, nonlinearities and nonreciprocity, cloaking and scattering, acoustics, optical nanocircuits and nanoantennas.